\documentclass[a4paper,11pt]{article}

\usepackage{jcappub} 

\usepackage[T1]{fontenc} 

\usepackage{amsmath,amssymb,amsfonts}
\usepackage{mathtools}
\usepackage{bm}
\usepackage{bbm}

\usepackage{graphicx}
\usepackage{xcolor}
\usepackage{caption}

\usepackage{comment}

\usepackage{natbib}
\usepackage{aas_macros}

\usepackage{hyperref}
\hypersetup{unicode=true}

\usepackage{amsthm}

\theoremstyle{definition}
\newtheorem{definition}{Definition}[section]

\theoremstyle{plain}

\theoremstyle{plain}
\newtheorem{proposition}{Proposition}[section]

\theoremstyle{plain}

\theoremstyle{remark}

\theoremstyle{remark}

\theoremstyle{remark}
\newtheorem{fact}{Fact}[section]

\def\pa#1#2{\dfrac{\partial #1}{\partial #2}}

\def\ta#1#2{\dfrac{\pd #1}{\pd #2}}

\def\pd{{\rm d}}

\def\deltadir{\delta_{\rm D}}

\title{A Measure-Theoretic Transport Formulation of Galaxy Evolution on the Galaxy Manifold: Geometric Constraints}

\author[a,b,1]{Tsutomu T. Takeuchi,\note{Corresponding author.}}

\affiliation[a]{Division of Particle and Astrophysical Science, Nagoya University, Furo-cho, Chikusa-ku, Nagoya, Aichi 464--8602, Japan}
\affiliation[b]{The Research Center for Statistical Machine
Learning, the Institute of Statistical Mathematics, 10--3 Midori-cho, Tachikawa, Tokyo 190--8562, Japan}

\emailAdd{tsutomu.takeuchi.ttt@gmail.com}

\abstract{We develop a measure-theoretic framework for galaxy evolution by describing galaxy populations as probability measures on a state space. 
In this formulation, galaxy evolution is represented as the time evolution of a measure $\nu_t$, whose dynamics are given by the sum of a continuous transport term and a jump operator. 
The transport term captures internal galaxy evolution, while the jump operator encodes discrete processes such as mergers and interactions, leading to a unified description as a reaction--transport system on the space of measures. 
We further endow the space of probability measures with the Wasserstein distance and impose a curvature--dimension condition CD$(K,\infty)$ to reveal the geometric structure of the evolution. 
Within this framework, the transport term is interpreted as a gradient flow of a free-energy functional, whereas the jump operator induces nonlinear rearrangements arising from many-body interactions. 
In the low-density limit, these interactions effectively reduce to two-body processes, yielding a closed dynamical system. 
A key consequence of this formulation is that galaxy evolution is not arbitrary but constrained by a variational structure, curvature bounds, and a hierarchy of interactions, such that admissible trajectories are restricted to those consistent with energy dissipation, geometric contractivity, and effective interaction closure. 
This theory thus provides a unified description of structure formation and galaxy evolution while specifying the class of allowed evolutionary paths. 
Moreover, the framework separates intrinsic galaxy dynamics from observational projection, treating observables independently as pushforwards of measures, thereby establishing a universal theoretical structure independent of specific observational methods or datasets. 
These results provide a new foundation for understanding galaxy evolution as a geometrically constrained reaction--transport process on the space of probability measures.
}

\begin{document}

\maketitle

\flushbottom

\section{Introduction}
\label{sec:introduction}

The formation and evolution of galaxies constitute one of the central problems in cosmology \citep[e.g.,][]{Mo2010GalaxyFormation, takeuchi2025physics}. 
In conventional theoretical approaches, galaxies are treated as individual objects, and their internal evolution and interactions are described based on their respective histories. 
However, observations always provide a finite ensemble of galaxies, and the quantities that are directly compared with theory are inherently statistical properties of such populations. 
From this perspective, it is essential to adopt a framework in which galaxy evolution is described not in terms of individual objects but as the evolution of a population distribution. 
This viewpoint provides the motivation for the concept of a galaxy manifold \citep[e.g.,][]{1973A&A....23..259B, 1992ASSL..178..337D, 2012MNRAS.419.1727C}. 

This perspective is consistent with the approach implicitly adopted in statistical analyses of large-scale structure. 
Galaxy distributions as observables are characterized by statistical quantities such as the two-point correlation function and the power spectrum, which are fundamentally functionals of the underlying distribution \citep[e.g.,][]{1980lssu.book.....P, Peacock1999Cosmology, martinez2002statistics, takeuchi2025physics}. 
By formulating the theoretical object directly as a measure, the correspondence with observables is established in a direct and structural manner. 
In this sense, describing galaxy populations as probability measures is not merely a formal extension but provides a natural framework that aligns the level of description between theory and observation \citep[e.g.,][]{takeuchi2026finite_sample_window}. 

The notion of a galaxy manifold has recently been employed heuristically in data-driven approaches, particularly in manifold learning and low-dimensional embedding methods \citep[e.g.,][]{2018A&A...617A..70S, 2023MNRAS.524.4976C}. 
While these methods are effective in extracting structures in the state space of galaxies from observational data, they are inherently algorithm-dependent, and the resulting structures depend sensitively on the choice of method and hyperparameters. 
Moreover, since the construction depends directly on observed quantities, the inferred manifold structure may change qualitatively when new data or different observational spaces are introduced \citep[cf.][]{ma2012manifold, Lee2007NonlinearDR}. 
In this sense, existing approaches remain at the level of phenomenological reconstruction and do not provide a theoretical framework that determines the intrinsic structure of the galaxy manifold \citep[see, e.g.,][]{2026Entrp..28..288T}. 
In contrast, in this work we define the galaxy manifold as a measurable space equipped with probability measures, and use the pushforward of measures as the fundamental operation to construct an intrinsic structure independent of observation and representation. 
In this formulation, the state space of galaxies is uniquely defined without reference to any algorithm, and mappings to observables are introduced externally as probability kernels or measurable maps. 
As a result, the projection onto observational data spaces is separated from the theoretical core, and intrinsic galaxy properties are described independently as structures on the space of measures. 
This separation ensures that the theoretical structure remains invariant under changes in observational methods or data availability, allowing the extraction of universal aspects of galaxy evolution. 

From the dynamical viewpoint, galaxy evolution involves two fundamentally distinct types of processes. 
One is continuous evolution, such as star formation and chemical enrichment \citep[e.g.,][]{1980FCPh....5..287T, Matteucci2012ChemicalEvolution}, and the other is discrete events such as mergers and interactions \citep[e.g.,][]{1978ApJ...220..390S, 1978ApJ...223L..59S, 1993MNRAS.262..627L}. 
In conventional models, these processes are often treated separately, whereas in reality they occur simultaneously and jointly shape the observed distribution. 
This motivates a unified treatment in which galaxy evolution is described as a coupling of continuous and jump processes. 
Such a decomposition corresponds to a standard structure in the theory of stochastic processes, and in this work we implement it rigorously as operators acting on the space of measures. 

The goal of this paper is to construct a rigorous measure-theoretic foundation for galaxy evolution by defining galaxy populations as probability measures on a state space and describing their time evolution. 
In this formulation, galaxy evolution is represented not as a collection of individual trajectories but as the evolution of a measure $\nu_t$. 
The time evolution is formulated operator-theoretically as the sum of continuous transport and discrete transitions, enabling a unified description of internal evolution and merger processes. 
A key point is that the transport component is not merely a dynamical flow but represents a rearrangement of mass distributions. 
Cosmological structure formation is described as the redistribution of matter under gravitational potentials, and this redistribution naturally corresponds to the framework of optimal transport \citep[e.g.,][]{2002Natur.417..260F, 2024PhRvD.109l3512N, Takeuchi2026OptimalTransport, takeuchi2026entropic_curvature_cosmo}. 
Thus, the time evolution of measures can be interpreted as a transport problem that moves mass from one distribution to another \citep{villani2003, villani2009optimal}. 
From this viewpoint, introducing a distance structure on the space of measures allows galaxy evolution to be reformulated as a problem in transport geometry. 

The methodology of this work proceeds by starting from a measure-theoretic formulation and systematically incorporating operator-theoretic and geometric structures to construct a closed theoretical framework for galaxy evolution. 
Specifically, we first define the galaxy manifold in terms of measurable spaces and probability measures, taking population distributions as the fundamental objects. 
We then formulate observation as a linear projection via probability kernels, thereby clearly separating latent states from observed distributions. 
The time evolution is introduced as an operator on the space of measures and decomposed into continuous transport and jump processes, providing a unified description of internal evolution and interactions. 
Furthermore, by introducing a distance structure, we equip the space of measures with transport geometry and extend the formulation to a variational framework. 
Through this stepwise construction, the structures of measure, operator, and geometry are systematically integrated, leading to a unified and progressively refined description of galaxy evolution. 

\begin{figure}[t]
\centering
\includegraphics[width=\textwidth]{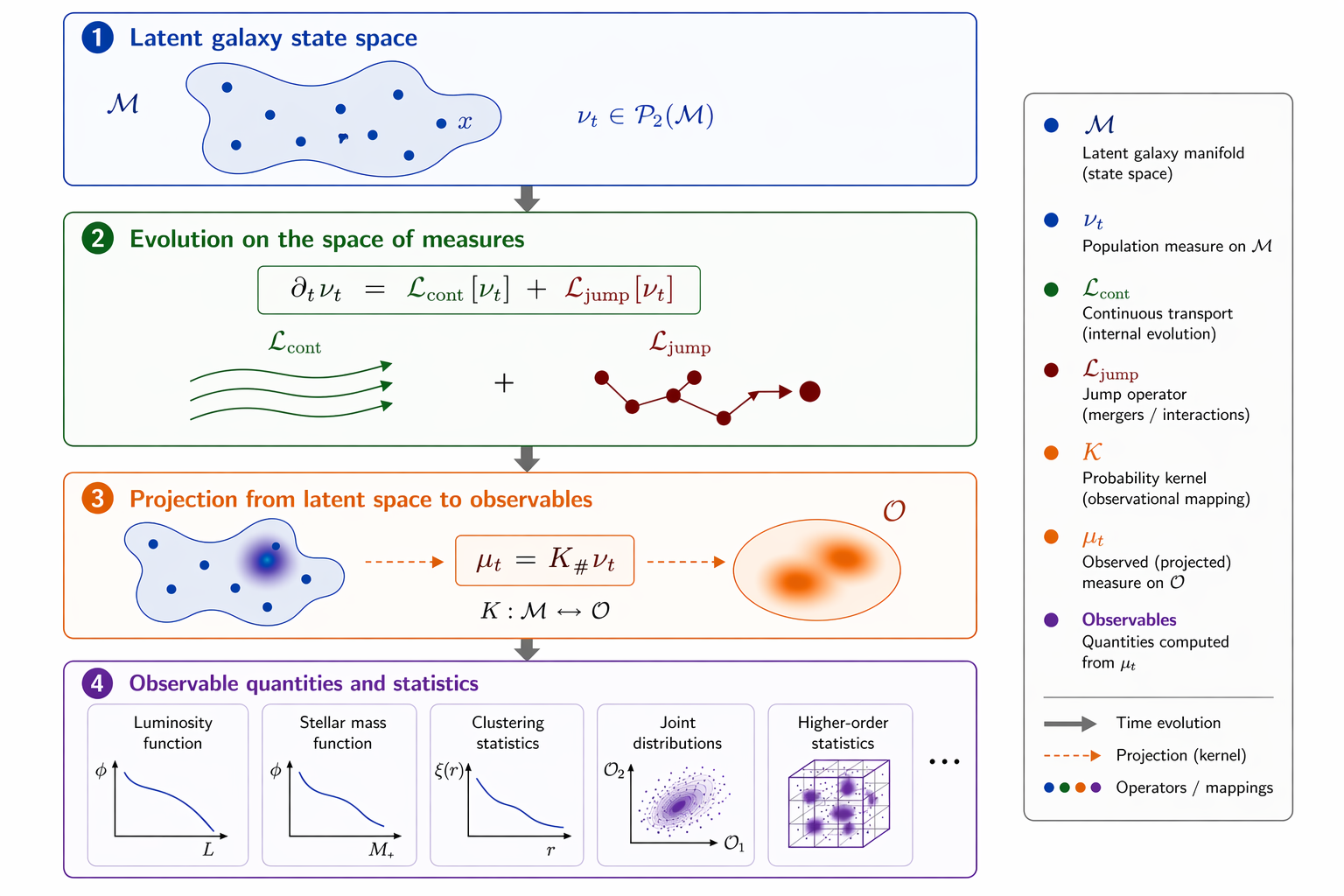}
\caption{
Schematic overview of the measure-theoretic formulation developed in this paper. Galaxy populations are represented as a probability measure $\nu_t$ on a latent state space $\mathcal{M}$. 
Its time evolution is described by the sum of continuous transport and jump processes. 
Observables are obtained via projection by a probability kernel, yielding an observed measure $\mu_t$. 
Statistical quantities are then constructed from the projected distribution. 
The framework cleanly separates intrinsic galaxy dynamics from their observational representation.
}
\label{fig:framework}
\end{figure}

In particular, the curvature--dimension condition CD$(K,\infty)$, introduced in the latter part of this work, provides a lower bound on the effective Ricci curvature of the space of measures and imposes geometric constraints on the stability and convergence of the evolution \citep{villani2009optimal, lott2009ricci, sturm2006geometryI, sturm2006geometryII}. 
Under this condition, continuous transport is not an arbitrary flow but is interpreted as a relaxation process governed by the geodesic convexity of th H-function\footnote{The H-function corresponds to a negative of entropy. 
Since we sometimes find a confusion in sign convention of the entropy, we only use H-function for consistency throughout this paper \citep[see, e.g.,][]{villani2009optimal}.}. 
Galaxy evolution is thus characterized as a reaction--transport system combining geometrically constrained transport with many-body interactions. 
The observational implications and direct applications of this measure-theoretic framework to real galaxy surveys will be presented in a companion paper \cite{takeuchi2026_merger_practical}.
The overall logical structure of the framework is summarized in Fig.~\ref{fig:framework}.

The structure of this paper is as follows. 
In Section~\ref{sec:measure_theoretic_formulation}, we define the galaxy manifold based on measure theory and introduce the fundamental structures of observation and time evolution. 
In Section~\ref{sec:pushforward_structure}, we formulate the representation structure as pushforwards of measures. 
In Section~\ref{sec:internal_evolution}, we describe internal galaxy evolution as a deterministic flow. 
In Section~\ref{sec:jump_process}, we construct merger processes rigorously as jump operators. 
In Section~\ref{sec:many_body_coalescence}, we introduce the hierarchical structure of many-body interactions and derive an effective two-body theory together with its convergence and scaling properties. 
In Section~\ref{sec:cd_transport}, we analyze measure evolution within the framework of transport geometry and the curvature--dimension condition CD$(K,\infty)$, and derive the unified structure of galaxy evolution and its consequences. 
The conclusions and outlook are presented in Section~\ref{sec:conclusion}. 
Basic definitions and supplementary material on measure theory, optimal transport, and curvature--dimension conditions are summarized in Appendix~\ref{app:measure_theory}, Appendix~\ref{app:ot}, and Appendix~\ref{app:cd}.

\section{Rigorous construction of the galaxy manifold based on measure theory}
\label{sec:measure_theoretic_formulation}

As discussed in Section~\ref{sec:introduction}, it is fundamentally important to formulate the foundational structure for describing galaxy evolution as a measure-theoretic framework that does not depend on any particular algorithm.
The purpose here is not to assume a manifold or a coordinate system a priori, but to define the evolving population of objects as a probability measure and then introduce geometric structure only when needed.
From this perspective, the time evolution of a galaxy population is understood not as a collection of trajectories of individual galaxies, but as a deformation of the distribution itself on the state space.
Accordingly, the fundamental object of the theory is not a point but a \emph{measure}, and the central task of this section is to describe its time evolution.
The structure of this section consists of the following three stages.
First, we define a galaxy population as a time-dependent probability measure on a state space.
Next, we formulate observation as a projection by a probability kernel, thereby separating the observed distribution from the latent distribution.
Finally, we introduce time evolution as an operator on the space of measures and define its decomposition into a continuous part and a jump part.
This construction provides a minimal and closed framework for describing galaxy evolution as the time evolution of a measure.

\subsection{State space and probability measure}
\label{subsec:state_space_and_measure}

We define the state space of galaxies as the measurable space
\begin{align}
	(\mathcal{M},\mathcal{F})
	\label{eq:measurable_state_space}
\end{align}
Here $\mathcal{F}$ is a $\sigma$-algebra on $\mathcal{M}$.
At this stage, it is important not to assume any topology or differentiable structure on $\mathcal{M}$, and to impose only measurability as the minimal structure.
Hereafter, let $\mathcal{P}(\mathcal{M})$ denote the set of all probability measures on $\mathcal{M}$.
A galaxy population is defined as a time-dependent measure
\begin{align}
	\nu_t\in\mathcal{P}(\mathcal{M})
	\label{eq:time_dependent_measure}
\end{align}
That is, for any $A\in\mathcal{F}$,
\begin{align}
	\nu_t(A)\in[0,1],\qquad \nu_t(\mathcal{M})=1
	\label{eq:probability_measure_condition}
\end{align}
holds.
In this case, $\nu_t$ gives the state distribution of the galaxy population at time $t$.

Furthermore, for the consistency of time evolution, we assume that for any bounded measurable function $f$,
\begin{align}
	t\longmapsto \int_{\mathcal{M}}f(z)\,\nu_t(\pd z)
	\label{eq:weak_continuity_measure}
\end{align}
is continuous.
This corresponds to the condition that the time evolution of the measure is continuous in the weak topology.

\begin{definition}
\label{def:galaxy_manifold}
In this work, we call the pair consisting of the state space $\mathcal{M}$ and the time-dependent probability measure $\nu_t$ on it,
\begin{align}
	(\mathcal{M},\nu_t)
	\label{eq:galaxy_manifold_pair}
\end{align}
the galaxy manifold.
\end{definition}
\medskip
\noindent
This definition is a purely measure-theoretic definition given prior to any geometric structure, and throughout this paper the primary meaning of the galaxy manifold lies in this pair.
The key point of this definition is that the galaxy manifold is regarded not as the geometry of a point set, but as the measure structure of a population distributed on a state space.
Therefore, galaxy evolution in this work is nothing other than the problem of asking how this measure $\nu_t$ is deformed with time.

\subsection{Observation space and probability kernel}
\label{subsec:observation_space_and_kernel}

Let the observation space be the measurable space
\begin{align}
	(\mathcal{X},\mathcal{B})
	\label{eq:observation_space}
\end{align}
In general, observables are not necessarily given deterministically as functions of latent states, and since they include measurement errors and selection effects, in this work we treat them as stochastic mappings.

The relation between latent states and observables is defined as a Markov kernel
\begin{align}
	K:\mathcal{M}\times\mathcal{B}\to[0,1]
	\label{eq:markov_kernel}
\end{align}
That is, it satisfies the following:
\begin{align}
	&\text{(i)}\quad B\mapsto K(B\mid z)\ \text{is a }\mathcal{B}\text{-probability measure},\notag\\
	&\text{(ii)}\quad z\mapsto K(B\mid z)\ \text{is }\mathcal{F}\text{-measurable}.
	\label{eq:markov_kernel_conditions}
\end{align}
In this case, the observed measure $\mu_t\in\mathcal{P}(\mathcal{X})$ is defined by
\begin{align}
	\mu_t(B)
	=
	\int_{\mathcal{M}}K(B\mid z)\,\nu_t(\pd z)
	\label{eq:observed_measure}
\end{align}
Therefore, observation is understood as the pushforward of the measure $\nu_t$.
Through this structure, observation and physical state are clearly separated (see Appendix~\ref{app:ot}).

\begin{proposition}
\label{prop:observation_duality}
For any bounded measurable function $\varphi:\mathcal{X}\to\mathbb{R}$,
\begin{align}
	\int_{\mathcal{X}}\varphi(x)\,\mu_t(\pd x)
	=
	\int_{\mathcal{M}}
	\left(
		\int_{\mathcal{X}}\varphi(x)\,K(\pd x\mid z)
	\right)
	\nu_t(\pd z)
	\label{eq:observation_duality}
\end{align}
holds.
\end{proposition}
\medskip
\noindent
This identity shows that observation is given as a linear action on the latent measure.
That is,
\begin{align}
	\nu_t\longmapsto\mu_t
	\label{eq:observation_linear_map}
\end{align}
is a linear map on measures, and observation is interpreted not as the dynamics itself but as a projection operation from the latent state distribution to the observed distribution.
This point becomes essential later when deriving the time evolution of observables.

\subsection{Definition of time evolution}
\label{subsec:time_evolution}

Galaxy evolution is defined in weak form by an operator $\mathcal{L}$.
That is, for any bounded measurable function $f$,
\begin{align}
	\ta{}{t}\int_{\mathcal{M}}f(z)\,\nu_t(\pd z)
	=
	\int_{\mathcal{M}}f(z)\,\mathcal{L}[\nu_t](\pd z)
	\label{eq:weak_evolution_equation}
\end{align}
when this holds, $\nu_t$ is said to satisfy the time evolution generated by $\mathcal{L}$.
This definition gives the time evolution of the measure as a dual action on the function space, and corresponds to the time derivative in the distributional sense (for weak derivatives and the distribution-theoretic framework, see Appendix~\ref{app:measure_theory}).

The operator $\mathcal{L}$ is decomposed as
\begin{align}
	\mathcal{L}
	=
	\mathcal{L}_{\mathrm{cont}}
	+
	\mathcal{L}_{\mathrm{jump}}
	\label{eq:generator_decomposition}
\end{align}
This decomposition corresponds to the standpoint of understanding galaxy evolution as a superposition of continuous deformations and discontinuous transitions.

\begin{definition}
\label{def:cont_jump_operators}
We call $\mathcal{L}_{\mathrm{cont}}$ the continuous transport operator and $\mathcal{L}_{\mathrm{jump}}$ the jump operator.
The former represents continuous changes of state, and the latter represents discrete transitions such as mergers.
\end{definition}
\medskip
\noindent
\begin{proposition}
\label{prop:evolution_nonlinearity}
Whereas the observation operator is linear, the full evolution operator $\mathcal{L}$ is generally nonlinear.
In particular, the nonlinearity originates from the jump term $\mathcal{L}_{\mathrm{jump}}$.
\end{proposition}
\medskip
\noindent
This contrast characterizes the structure of the present theory.
That is, whereas observation is a linear projection, physical evolution is generally nonlinear, and this is also the fundamental reason why it is difficult to write the evolution of observables as a closed system.

\subsection{Continuous transport term}
\label{subsec:continuous_transport}

Consider a measurable vector field $v(z,t)$ on $\mathcal{M}$, and for any $f\in C_c^1(\mathcal{M})$, define
\begin{align}
	\int_{\mathcal{M}}f(z)\,\mathcal{L}_{\mathrm{cont}}[\nu_t](\pd z)
	=
	\int_{\mathcal{M}}\nabla f(z)\cdot v(z,t)\,\nu_t(\pd z)
	\label{eq:continuous_transport_weak}
\end{align}
This expresses that the measure is transported by a flow on the state space.
Furthermore, if a density $\rho(z,t)$ exists and
\begin{align}
	\nu_t(\pd z)=\rho(z,t)\pd z
	\label{eq:density_representation}
\end{align}
can be written, then in weak form $\rho$ satisfies
\begin{align}
	\pa{\rho}{t}
	+
	\nabla\cdot(\rho v)
	=
	0
	\label{eq:continuity_equation}
\end{align}
This equation is a continuity equation of mass-conserving type and describes continuous processes such as star formation and chemical evolution.

\begin{proposition}
\label{prop:cont_operator_linearity}
The continuous transport operator $\mathcal{L}_{\mathrm{cont}}$ is linear with respect to the measure.
That is, for any probability measures $\nu,\eta$ and nonnegative constants $a,b$,
\begin{align}
	\mathcal{L}_{\mathrm{cont}}[a\nu+b\eta]
	=
	a\mathcal{L}_{\mathrm{cont}}[\nu]
	+
	b\mathcal{L}_{\mathrm{cont}}[\eta]
	\label{eq:cont_operator_linearity}
\end{align}
holds.
\end{proposition}
\medskip
\noindent
Therefore, the nonlinearity is not intrinsically tied to the continuous transport itself, but rather to the jump process arising from two-body interactions.

\subsection{Jump operator}
\label{subsec:jump_operator}

Define the merger kernel as a measurable function
\begin{align}
	K(z_1,z_2\to z;t)\ge0
	\label{eq:coagulation_kernel}
\end{align}
and assume that the following integral is finite:
\begin{align}
	\int_{\mathcal{M}}\int_{\mathcal{M}}K(z_1,z_2\to z;t)\,\nu(\pd z_1)\nu(\pd z_2)<\infty
	\label{eq:coagulation_kernel_integrability}
\end{align}
We further assume the symmetry
\begin{align}
	K(z_1,z_2\to z;t)=K(z_2,z_1\to z;t)
	\label{eq:coagulation_kernel_symmetry}
\end{align}
This corresponds to the fact that the merger process is an unordered two-body reaction.

Define the destruction rate by
\begin{align}
	\kappa(z,z';t)
	=
	\int_{\mathcal{M}}K(z,z'\to\tilde{z};t)\,\pd \tilde{z}
	\label{eq:destruction_rate}
\end{align}

Then the jump operator is defined for any $f$ by
\begin{align}
	&\int_{\mathcal{M}}f(z)\,\mathcal{L}_{\mathrm{jump}}[\nu_t](\pd z) \notag \\
	&\qquad =
	\frac{1}{2}
	\int_{\mathcal{M}}\int_{\mathcal{M}}
	\left[
	\int_{\mathcal{M}}f(z)K(z_1,z_2\to z;t)\,\pd z
	-
	f(z_1)
	-
	f(z_2)
	\right]
	\nu_t(\pd z_1)\nu_t(\pd z_2)
	\label{eq:jump_operator}
\end{align}
Rewriting this as
\begin{align}
	&\int_{\mathcal{M}}f(z)\,\mathcal{L}_{\mathrm{jump}}[\nu_t](\pd z) \notag \\
	&\qquad =
	\frac{1}{2}
	\int_{\mathcal{M}}\int_{\mathcal{M}}
	\Bigl[
		\langle f, K(z_1,z_2\to \cdot;t)\rangle
		- f(z_1) - f(z_2)
	\Bigr]
	\nu_t(\pd z_1)\nu_t(\pd z_2)
	\label{eq:jump_operator2}
\end{align}
this structure is a nonlinear operator based on two-body interactions and is a natural generalization of the Smoluchowski-type coagulation process.

\begin{proposition}
\label{prop:jump_quadratic_nonlinearity}
The jump operator $\mathcal{L}_{\mathrm{jump}}$ is a quadratic nonlinear operator with respect to the measure.
That is, its weak form depends on the product measure $\nu_t\otimes\nu_t$, and has the structure
\begin{align}
	\mathcal{L}_{\mathrm{jump}}[\nu_t]
	\sim
	\nu_t\otimes\nu_t
	\label{eq:jump_quadratic_structure}
\end{align}
\end{proposition}
\medskip
\noindent
This means that mergers are intrinsically two-body processes and represent a dynamical hierarchy different from the transport of a single galaxy.
Moreover, while Eq.~(\ref{eq:jump_operator}) is written as the difference between a gain term and a loss term, Eq.~(\ref{eq:jump_operator2}) is an equivalent expression that makes explicit the bilinear structure of the operator.
The two are equivalent, but the latter is more suitable for operator-theoretic discussion.

\subsection{Conserved quantities}
\label{subsec:conserved_quantities}

When a measurable function $q:\mathcal{M}\to\mathbb{R}$ satisfies
\begin{align}
	q(z)
	=
	q(z_1)+q(z_2)
	\label{eq:additive_conserved_quantity}
\end{align}
then
\begin{align}
	\ta{}{t}
	\int_{\mathcal{M}}q(z)\,\nu_t(\pd z)
	=
	0
	\label{eq:conserved_moment}
\end{align}
holds.
This means that the conserved quantity is invariant under the jump process.

\begin{proposition}
\label{prop:moment_conservation}
Under the above condition, the moment corresponding to $q$,
\begin{align}
	M_q(t)
	=
	\int_{\mathcal{M}}q(z)\,\nu_t(\pd z)
	\label{eq:moment_definition}
\end{align}
is independent of time.
\end{proposition}
\medskip
\noindent
This proposition shows that even if the jump operator does not conserve the particle number itself, it can nevertheless preserve invariance for additive conserved quantities.
In the context of galaxy mergers, mass, certain total metal content, or suitably defined conservative mark quantities belong to this class.
Therefore, the present framework simultaneously allows a decrease in number density and the preservation of conserved quantities.
This point is important in the physical interpretation of coagulation equations.

\subsection{Relation to finite samples}
\label{subsec:finite_sample_relation}

Observed data are given as a finite set
\begin{align}
	X_N=\{x_1,\dots,x_N\}
	\label{eq:finite_sample}
\end{align}
and the corresponding empirical measure is
\begin{align}
	\mu_N
	=
	\frac{1}{N}\sum_{i=1}^N\deltadir(x_i)
	\label{eq:empirical_measure}
\end{align}
Here $\deltadir(x)$ denotes the Dirac delta function.

$\mu_N$ is an empirical approximation of $\mu_t$, and we assume the weak convergence
\begin{align}
	\mu_N \Rightarrow \mu_t
	\label{eq:empirical_weak_convergence}
\end{align}
(for the definition of weak convergence, see Appendix~\ref{app:measure_theory}).
This relation gives the correspondence between finite data and the measure-theoretic limit.

\begin{proposition}
\label{prop:empirical_measure_lift}
The empirical measure $\mu_N$ is the standard construction that lifts the finite point set $X_N$ to a measure-theoretic object.
Accordingly, the present theory provides a framework that starts from finitely many observed data points while describing their limit on the space of measures.
\end{proposition}
\medskip
\noindent
From this viewpoint, finite samples and continuous distributions are not opposing concepts, but descriptions at different levels of the same theory.
A finite sample is embedded into measure theory as an empirical measure, and its limit appears as the observed measure $\mu_t$.
Therefore, the measure-theoretic formulation of the galaxy manifold introduced in this work incorporates from the outset its connection to finite-data analysis.

\subsection{Geometric structure}
\label{subsec:geometric_structure}

In the construction up to this point, the only structure required of $\mathcal{M}$ is that of a measurable space.
Therefore, the primary content of the galaxy manifold lies strictly in the state space and the measure on it.
However, to interpret the time evolution of measures geometrically, it is useful to introduce a metric structure on $\mathcal{M}$.
When geometric comparison of measures becomes necessary, we may endow $\mathcal{M}$ with a compatible metric.
We therefore let $(\mathcal{M},d)$ be a Polish space.
Furthermore,
\begin{align}
	\mathcal{P}_2(\mathcal{M})
	=
	\left\{
		\nu\in\mathcal{P}(\mathcal{M})
		\ \middle|\
		\int_{\mathcal{M}}d(z,z_0)^2\,\nu(\pd z)<\infty
		\ \text{for some }z_0\in\mathcal{M}
	\right\}
	\label{eq:p2_space}
\end{align}
is defined as the set of probability measures with finite second moment (for the properties of this space and the definition of the Wasserstein distance, see Appendix~\ref{app:ot}).
By the Polish property, this condition does not depend on the choice of the reference point $z_0$.

\begin{definition}
\label{def:metric_galaxy_manifold}
We call $(\mathcal{M},d,\nu_t)$ a metric galaxy manifold.
This is obtained by adding a metric structure to the previously introduced measure-theoretic galaxy manifold $(\mathcal{M},\nu_t)$.
\end{definition}
\medskip
\noindent
If we then assume
\begin{align}
	\nu_t\in\mathcal{P}_2(\mathcal{M})
	\label{eq:nu_in_p2}
\end{align}
a natural geometric structure induced by optimal transport is defined on the space of measures \citep{villani2009optimal}.
In this work, the details will be introduced later to the extent necessary, but the important point here is that geometry is not the starting point, but a secondary structure added to the measure structure.

The galaxy manifold defined in this section is first defined as a pair of a measurable space and a probability measure, and is extended to a geometric object by adding a metric structure when necessary.
That is, the essence of the galaxy manifold lies not in a geometrical space given from the outset, but in the measure structure of galaxy populations and their dynamics.
In Section~\ref{sec:cd_transport} of this paper, we will see that geometry appears as an auxiliary yet powerful structure for describing the comparison and transport of measures.

\section{Manifold representation as the pushforward of measures}
\label{sec:pushforward_structure}

In this section, we formulate low-dimensional representations as the pushforward of measurable mappings acting on measures defined on the observation space.
In Section~\ref{sec:measure_theoretic_formulation}, the observation space $(\mathcal{X},\mathcal{B})$ and the observed measure $\mu_t\in\mathcal{P}(\mathcal{X})$ are defined by Eq.~(\ref{eq:observation_space}) and Eq.~(\ref{eq:observed_measure}), respectively.
In this section, we consider the operation of constructing a new measure by applying a measurable mapping to this observed measure.
This operation does not define the galaxy manifold itself, but instead provides a secondary representation defined on the observed measure $\mu_t$.
Accordingly, the purpose of this section is not to assume a manifold, but to clarify the structure that appears as the image of a measure.
In the following, we first revisit the relation between finite samples and empirical measures introduced in Section~\ref{sec:measure_theoretic_formulation}, then give the definition and basic properties of the pushforward.
We then summarize the structural properties of the pushforward and finally organize them as a hierarchical structure of measures.

\subsection{Finite sample approximation}
\label{subsec:finite_sample_pushforward}

The basic construction of finite samples and empirical measures has already been introduced in Section~\ref{sec:measure_theoretic_formulation}, Subsection~\ref{subsec:finite_sample_relation}.
That is, the finite set $X_N$ and the empirical measure $\mu_N$ are defined by Eq.~(\ref{eq:finite_sample}) and Eq.~(\ref{eq:empirical_measure}), respectively, and the weak convergence to the observed measure $\mu_t$ is given by Eq.~(\ref{eq:empirical_weak_convergence}) (see Appendix~\ref{app:measure_theory}).
Therefore, in this section we do not redefine them, but use them as finite-sample approximations of the observed measure $\mu_t$.

In this sense, a finite set of observed galaxy data is embedded within the same framework as an empirical realization of the observed measure $\mu_t$.
Moreover, as stated in Section~\ref{sec:measure_theoretic_formulation}, the empirical measure is the standard construction that lifts a finite point set into a measure-theoretic object.
Therefore, the discussion of pushforward that follows applies equally to the continuous observed measure $\mu_t$ and to its finite-sample approximation $\mu_N$.

\begin{proposition}
\label{prop:empirical_average_pushforward_section}
For any bounded measurable function $f:\mathcal{X}\to\mathbb{R}$,
\begin{align}
	\int_{\mathcal{X}} f(x)\,\mu_N(\pd x)
	=
	\frac{1}{N}\sum_{i=1}^N f(x_i)
	\label{eq:empirical_average_formula}
\end{align}
holds.
\end{proposition}
\medskip
\noindent
\begin{proof}
Using Eq.~(\ref{eq:empirical_measure}),
\begin{align}
	\int_{\mathcal{X}} f(x)\,\mu_N(\pd x)
	=
	\int_{\mathcal{X}} f(x)\,
	\left(
	\frac{1}{N}\sum_{i=1}^N\deltadir(x_i)
	\right)(\pd x)
	=
	\frac{1}{N}\sum_{i=1}^N f(x_i)
\end{align}
follows.
\end{proof}
\medskip
\noindent
This result shows that the empirical measure is a linear functional that yields the sample average.
Therefore, any representation statistic defined for a finite sample can be rewritten as an integral with respect to the empirical measure.
This unifies representation-learning operations on finite data with pushforward operations on the space of measures in the same formalism.

\subsection{Definition of pushforward}
\label{subsec:pushforward_definition}

Consider measurable spaces $(\mathcal{X},\mathcal{B})$ and $(\mathcal{Z},\mathcal{C})$, and a measurable mapping
\begin{align}
	\Phi:\mathcal{X}\to\mathcal{Z}
	\label{eq:representation_map}
\end{align}
Here $\Phi$ is a mapping defined on measures over the observation space, and in the context of the galaxy manifold it is interpreted as an operation that provides a representation of the observed distribution.
That is, it is a mapping that assigns a new measure as the image of the observed measure $\mu_t$ constructed in Section~\ref{sec:measure_theoretic_formulation}.

\begin{definition}
\label{def:pushforward_measure}
For a measure $\mu\in\mathcal{P}(\mathcal{X})$,
\begin{align}
	\nu=\Phi_{\#}\mu
	\label{eq:pushforward_measure_symbol}
\end{align}
is called the pushforward measure and is defined by
\begin{align}
	\nu(A)=\mu(\Phi^{-1}(A))
	\label{eq:pushforward_measure_definition}
\end{align}
\end{definition}
\medskip
\noindent
\begin{proposition}
\label{prop:pushforward_probability_measure}
$\Phi_{\#}\mu\in\mathcal{P}(\mathcal{Z})$.
\end{proposition}
\medskip
\noindent
\begin{proof}
By the measurability of $\Phi$, for any $A\in\mathcal{C}$ we have $\Phi^{-1}(A)\in\mathcal{B}$, so Eq.~(\ref{eq:pushforward_measure_definition}) defines a measure on $\mathcal{Z}$.
Furthermore,
\begin{align}
	\Phi_{\#}\mu(\mathcal{Z})
	=
	\mu(\Phi^{-1}(\mathcal{Z}))
	=
	\mu(\mathcal{X})
	=
	1
\end{align}
and hence $\Phi_{\#}\mu$ is a probability measure.
\end{proof}
\medskip
\noindent
Through this operation, a measure is transported to another space via a measurable mapping.
Thus, a low-dimensional representation is understood not as a coordinate transformation of point sets, but as a pushforward measure defined on the observed measure.

\begin{proposition}
\label{prop:pushforward_integral_formula}
For any bounded measurable function $f:\mathcal{Z}\to\mathbb{R}$,
\begin{align}
	\int_{\mathcal{Z}}f(z)\,\nu(\pd z)
	=
	\int_{\mathcal{X}}f\bigl(\Phi(x)\bigr)\,\mu(\pd x)
	\label{eq:pushforward_integral_formula}
\end{align}
holds.
\end{proposition}
\medskip
\noindent
\begin{proof}
First, for the indicator function $f=\mathbf{1}_A$,
\begin{align}
	\int_{\mathcal{Z}}\mathbf{1}_A(z)\,\nu(\pd z)
	=
	\nu(A)
	=
	\mu(\Phi^{-1}(A))
	=
	\int_{\mathcal{X}}\mathbf{1}_A(\Phi(x))\,\mu(\pd x)
\end{align}
holds.
By linearity this extends to simple functions, and further to bounded measurable functions by the monotone convergence theorem or the bounded convergence theorem.
\end{proof}
\medskip
\noindent
This relation shows that the pushforward is an integral transformation defined via function composition (see Appendix~\ref{app:ot} for the integral transformation formula).
Therefore, statistical quantities on the representation space are uniquely determined as pullbacks of functions on the observed measure.
As in the dual representation Eq.~(\ref{eq:observation_duality}) of the observation kernel $K$ in Section~\ref{sec:measure_theoretic_formulation}, the image of the measure and the pullback of functions form a dual relationship here as well.
In this sense, the pushforward is positioned as a second-stage projection on the observation space following the projection from latent states to observation via the observation kernel (Appendix~\ref{app:ot}).

\subsection{Structure of pushforward}
\label{subsec:pushforward_structure_properties}

The pushforward is not merely a mapping of measures but carries several fundamental structures.

\begin{fact}
\label{fact:formal_density_pushforward}
When a density exists, formally
\begin{align}
	\nu(z)
	=
	\int_{\mathcal{X}}
	\deltadir(z-\Phi(x))\,\mu(\pd x)
	\label{eq:formal_pushforward_density}
\end{align}
can be written.
\end{fact}
\medskip
\noindent
This expression shows that the representation measure is constructed as a superposition of the observed measure.
That is, each observed point $x$ is mapped to a point $\Phi(x)$ in the representation space, and the overall distribution $\nu$ is obtained as their aggregate.
In the case of finite samples, substituting Eq.~(\ref{eq:empirical_measure}) into Eq.~(\ref{eq:pushforward_measure_symbol}) yields
\begin{align}
	\Phi_{\#}\mu_N
	=
	\frac{1}{N}\sum_{i=1}^N\deltadir\bigl(\Phi(x_i)\bigr)
	\label{eq:empirical_pushforward}
\end{align}
Therefore, the empirical distribution of embeddings for point cloud data is rigorously identified as the pushforward of the empirical measure.

\begin{fact}
\label{fact:noninjective_fibers}
If $\Phi$ is not injective, then $\Phi^{-1}(z)$ contains multiple elements.
\end{fact}
\medskip
\noindent
\begin{proposition}
\label{prop:pushforward_noninvertible}
If $\Phi$ is not injective, then $\mu$ is not uniquely determined from $\Phi_{\#}\mu$.
\end{proposition}
\medskip
\noindent
\begin{proof}
Suppose that two distinct points $x_1\neq x_2$ satisfy $\Phi(x_1)=\Phi(x_2)$.
Then
\begin{align}
	\mu_1=\deltadir(x_1),\qquad \mu_2=\deltadir(x_2)
\end{align}
are distinct measures, but
\begin{align}
	\Phi_{\#}\mu_1
	=
	\deltadir(\Phi(x_1))
	=
	\deltadir(\Phi(x_2))
	=
	\Phi_{\#}\mu_2
\end{align}
holds.
Hence $\mu$ cannot be uniquely recovered from $\Phi_{\#}\mu$.
\end{proof}
\medskip
\noindent
Therefore, the pushforward is generally not invertible.
This non-invertibility implies that the representation compresses information of the original measure.
In the context of the galaxy manifold, low-dimensional representations obtained from the observed measure $\mu_t$ preserve certain aspects of the distribution while potentially losing detailed discriminative information in the original observation space.
In this sense, pushforward representations are intrinsically information compression.

\begin{proposition}
\label{prop:pullback_duality_operator}
The mapping
\begin{align}
	T_{\Phi}f=f\circ\Phi
	\label{eq:pullback_operator}
\end{align}
is a linear operator and satisfies
\begin{align}
	\int_{\mathcal{Z}} f(z)\,\nu(\pd z)
	=
	\int_{\mathcal{X}} T_{\Phi}f(x)\,\mu(\pd x)
	\label{eq:pullback_duality}
\end{align}
\end{proposition}
\medskip
\noindent
\begin{proof}
The linearity $T_{\Phi}(af+bg)=aT_{\Phi}f+bT_{\Phi}g$ follows immediately.
Equation~(\ref{eq:pullback_duality}) is obtained by substituting $T_{\Phi}f=f\circ\Phi$ into Eq.~(\ref{eq:pushforward_integral_formula}).
\end{proof}
\medskip
\noindent
This relation shows that the pushforward is dual to the pullback of functions.
As in Section~\ref{sec:measure_theoretic_formulation}, where observation appears as a linear action on the latent measure, here again the representation measure is described via a linear operator $T_{\Phi}$ on function space.
Therefore, the representation considered in this section is an example of a linear projection constructed on the observed measure.

\begin{definition}
\label{def:pushforward_equivalence_relation}
Define the equivalence relation
\begin{align}
	x\sim x'
	\iff
	\Phi(x)=\Phi(x')
	\label{eq:pushforward_equivalence_relation}
\end{align}
\end{definition}

\begin{fact}
\label{fact:quotient_measure_interpretation}
$\nu$ can be interpreted as a measure on the quotient space $\mathcal{X}/\sim$.
\end{fact}
\medskip
\noindent
Thus, the pushforward is understood as a projection onto equivalence classes.
Therefore, a representation does not retain each point in the observation space individually, but corresponds to viewing the quotient structure induced by an indistinguishability relation as a measure.
This perspective clarifies that dimensionality reduction and feature extraction are not mere coordinate transformations but operations involving identification of information.

From the above,
\begin{align}
	\mu_t \longmapsto \nu=\Phi_{\#}\mu_t
	\label{eq:measure_hierarchy_pushforward}
\end{align}
is defined.
Furthermore, using the latent measure $\nu_t$, the observation kernel $K$, and the observed measure $\mu_t$ introduced in Section~\ref{sec:measure_theoretic_formulation},
\begin{align}
	\nu_t
	\longmapsto
	\mu_t
	\longmapsto
	\Phi_{\#}\mu_t
	\label{eq:latent_observed_representation_hierarchy}
\end{align}
a hierarchy is obtained.
Here the first arrow is the projection by the observation kernel given by Eq.~(\ref{eq:observed_measure}), and the second arrow is the pushforward by the measurable mapping $\Phi$ introduced in this section.
Therefore, the representation structure in this section is clearly positioned as a secondary construction built on top of Section~\ref{sec:measure_theoretic_formulation}.

\begin{definition}
\label{def:pushforward_measure_hierarchy}
We call $(\mu_t,\Phi_{\#}\mu_t)$ the hierarchy of measures induced by pushforward.
\end{definition}
\medskip
\noindent
This structure is closed under measures and measurable mappings.
In the construction of the galaxy manifold, this hierarchy corresponds to the stage of representation obtained from the observed measure $\mu_t$, and is distinguished from the measure on the latent state space.
Therefore, the results of this section do not construct the manifold itself, but provide general properties of the representation structure that arises from the observed measure.
More precisely, the primary content of the galaxy manifold lies in the measure-theoretic construction given in Definition~\ref{def:galaxy_manifold}, and the pushforward representation in this section constitutes an observational and representational layer built upon it.
By maintaining this distinction, one can discuss \emph{the manifold itself and the low-dimensional representations induced from the manifold without confusion}.

\section{Deterministic flow of internal galaxy evolution}
\label{sec:internal_evolution}

From the previous sections, galaxy evolution has been formulated as the time evolution of a measure $\nu_t\in\mathcal{P}(\mathcal{M})$ on the state space.
In particular, in Section~\ref{sec:measure_theoretic_formulation}, the weak form of time evolution is given by Eq.~(\ref{eq:weak_evolution_equation}), the continuous transport term by Eq.~(\ref{eq:continuous_transport_weak}), and, under a density representation, the continuity equation by Eq.~(\ref{eq:continuity_equation}).
In this section, we concretize the part corresponding to the continuous transport term within this general structure as a deterministic flow representing the internal evolution of galaxies.
This structure provides, on the galaxy manifold, a deterministic evolution described by an ordinary differential equation for the state variables.

In the following, we first define the time evolution as a measurable semiflow, and then give its realization in terms of ordinary differential equations.
We then organize the time evolution of measures via pushforward and the continuity equation in the density representation as concrete instances of the general framework of Section~\ref{sec:measure_theoretic_formulation}, and incorporate the chemical evolution equations into this framework.
Finally, we formulate spectral evolution as a measurable mapping and express observables as pushforwards of measures.

\subsection{Internal state space and deterministic flow}
\label{subsec:internal_state_and_flow}

Let $(\mathcal{M}_{\mathrm{int}},\mathcal{F}_{\mathrm{int}})$ be a measurable space describing internal states.
Each point $\theta\in\mathcal{M}_{\mathrm{int}}$ represents the internal state of a galaxy.
In this section, we first define time evolution on this space as an abstract semiflow, and then realize it in terms of an ordinary differential equation.

\subsubsection{Definition of measurable semiflow}
\label{subsubsec:measurable_semiflow}

\begin{definition}
\label{def:measurable_semiflow}
A family of measurable mappings
\begin{align}
	\phi_t:\mathcal{M}_{\mathrm{int}}\to\mathcal{M}_{\mathrm{int}},
	\qquad t\ge0
	\label{eq:measurable_semiflow}
\end{align}
is called a deterministic semiflow if
\begin{align}
	\phi_0=\mathrm{Id},
	\qquad
	\phi_{t+s}=\phi_t\circ\phi_s
	\label{eq:semiflow_property}
\end{align}
holds for any $t,s\ge0$.
Here $\mathrm{Id}$ denotes the identity map.
\end{definition}
\medskip
\noindent
This definition means that time evolution is given as a semigroup of measurable mappings.
That is, each $\theta$ is uniquely mapped to $\phi_t(\theta)$ as time evolves.
In Section~\ref{sec:measure_theoretic_formulation}, time evolution was described as a generator acting on measures, whereas here we first describe the continuous part as the time evolution of points and then lift it to measures.

\subsubsection{Realization by ordinary differential equations}
\label{subsubsec:ode_realization}

Hereafter, we regard $\mathcal{M}_{\mathrm{int}}$ as a measurable subset of $\mathbb{R}^d$, and introduce a measurable mapping
\begin{align}
	F:\mathcal{M}_{\mathrm{int}}\times[0,\infty)\to\mathbb{R}^{d}
	\label{eq:internal_vector_field}
\end{align}

\begin{definition}
\label{def:internal_ode_semiflow}
If the solution of the ordinary differential equation
\begin{align}
	\ta{\theta}{t}=F(\theta,t)
	\label{eq:internal_ode}
\end{align}
exists and is unique, then the family of mappings $\{\phi_t\}_{t\ge0}$ defined by this solution is called the semiflow of internal evolution (see Appendix~\ref{app:measure_theory} for existence and uniqueness conditions).
That is,
\begin{align}
	\phi_t(\theta_0)=\theta(t)
	\label{eq:semiflow_from_ode}
\end{align}
\end{definition}
\medskip
\noindent
In this case, $\{\phi_t\}_{t\ge0}$ forms a semiflow in the sense of Definition~\ref{def:measurable_semiflow}.
That is, time evolution defined by an ordinary differential equation provides a concrete realization of the abstract semiflow.
The deterministic internal evolution considered in this section refers, in this sense, to the flow given as the motion of a point $\theta$.

\subsection{Time evolution of measures and densities}
\label{subsec:measure_and_density_evolution}

The semiflow $\{\phi_t\}_{t\ge0}$ of internal evolution induces the time evolution of measures.
In Section~\ref{sec:pushforward_structure}, the general theory of pushforward is given, in particular Definition~\ref{def:pushforward_measure} and Eq.~(\ref{eq:pushforward_integral_formula}).
In this subsection, we apply it to the semiflow $\phi_t$ on the internal state space and concretize the continuous transport term in Section~\ref{sec:measure_theoretic_formulation}.

\subsubsection{Measure evolution by pushforward}
\label{subsubsec:measure_evolution_by_pushforward}

\begin{proposition}
\label{prop:internal_measure_pushforward}
For an initial measure $\nu_0\in\mathcal{P}(\mathcal{M}_{\mathrm{int}})$,
\begin{align}
	\nu_t=(\phi_t)_{\#}\nu_0
	\label{eq:internal_measure_pushforward}
\end{align}
defines $\nu_t$ as an element of $\mathcal{P}(\mathcal{M}_{\mathrm{int}})$.
\end{proposition}
\medskip
\noindent
\begin{proof}
This follows immediately by applying Proposition~\ref{prop:pushforward_probability_measure} of Section~\ref{sec:pushforward_structure} with $\Phi=\phi_t$.
\end{proof}
\medskip
\noindent
Through this relation, internal evolution is described as a deterministic pushforward acting on measures.
That is, the time evolution of the measure is given not as a collection of trajectories but as the image measure under the mapping $\phi_t$.
This corresponds to the case in the general framework of Section~\ref{sec:measure_theoretic_formulation} where only continuous transport governs the evolution.

\subsubsection{Density representation and continuity equation}
\label{subsubsec:density_and_continuity_equation}

Assuming absolute continuity with respect to a reference measure $\lambda$,
\begin{align}
	\nu_t(\pd \theta)=\rho(\theta,t)\pd \theta
	\label{eq:internal_density_representation}
\end{align}

First, we define the time evolution of the measure in weak form.

\begin{definition}
\label{def:weak_time_differentiability_internal}
A family of measures $\{\nu_t\}_{t\ge0}$ is said to be weakly differentiable in time if, for any $f\in C_c^1(\mathcal{M}_{\mathrm{int}})$,
\begin{align}
	t\longmapsto
	\int_{\mathcal{M}_{\mathrm{int}}}
	f(\theta)\,\nu_t(\pd \theta)
	\label{eq:weak_time_differentiability_map}
\end{align}
is differentiable, and its derivative is given by
\begin{align}
	\ta{}{t}
	\int_{\mathcal{M}_{\mathrm{int}}}
	f(\theta)\,\nu_t(\pd \theta)
	\label{eq:weak_time_derivative_internal}
\end{align}
\end{definition}
\medskip
\noindent
This definition gives the time derivative of the measure through integration against test functions and corresponds to the distributional time derivative.
Equation~(\ref{eq:weak_evolution_equation}) in Section~\ref{sec:measure_theoretic_formulation} provides the general framework, and the present subsection constitutes its concrete realization.

\begin{proposition}
\label{prop:weak_form_internal_flow}
For the measure defined by the semiflow $\phi_t$,
\begin{align}
	\nu_t=(\phi_t)_{\#}\nu_0
	\label{eq:internal_measure_pushforward_repeat}
\end{align}
the following holds for any $f\in C_c^1(\mathcal{M}_{\mathrm{int}})$:
\begin{align}
	\ta{}{t}
	\int_{\mathcal{M}_{\mathrm{int}}}
	f(\theta)\,\nu_t(\pd \theta)
	=
	\int_{\mathcal{M}_{\mathrm{int}}}
	\nabla f(\theta)\cdot F(\theta,t)\,\nu_t(\pd \theta)
	\label{eq:weak_form_internal_flow}
\end{align}
\end{proposition}
\medskip
\noindent
\begin{proof}
Applying Eq.~(\ref{eq:pushforward_integral_formula}) of Section~\ref{sec:pushforward_structure} with $\Phi=\phi_t$, we obtain
\begin{align}
	\int_{\mathcal{M}_{\mathrm{int}}} f(\theta)\,\nu_t(\pd \theta)
	=
	\int_{\mathcal{M}_{\mathrm{int}}}
	f\bigl(\phi_t(\theta_0)\bigr)\,\nu_0(\pd \theta_0)
	\label{eq:proof_internal_pushforward}
\end{align}
Differentiating both sides with respect to $t$ and using the chain rule
\begin{align}
	\ta{}{t}f(\phi_t(\theta_0))
	=
	\nabla f(\phi_t(\theta_0))\cdot F(\phi_t(\theta_0),t)
	\label{eq:chain_rule_internal_flow}
\end{align}
yields the result.
\end{proof}
\medskip
\noindent
Equation~(\ref{eq:weak_form_internal_flow}) is nothing but the specialization of the general continuous transport operator in Eq.~(\ref{eq:continuous_transport_weak}) of Section~\ref{sec:measure_theoretic_formulation}, where the velocity field $v$ is concretized as the vector field $F$ on the internal state space.
Thus, the present construction does not introduce a new principle, but applies the general theory of continuous transport to internal galaxy evolution.

Furthermore, under the density representation,
\begin{align}
	\ta{}{t}
	\int_{\mathcal{M}_{\mathrm{int}}}
	f(\theta)\rho(\theta,t)\pd \theta
	=
	\int_{\mathcal{M}_{\mathrm{int}}}
	\nabla f(\theta)\cdot F(\theta,t)\rho(\theta,t)\pd \theta
	\label{eq:weak_form_density_internal}
\end{align}
holds.

\begin{proposition}
\label{prop:internal_continuity_equation}
The density $\rho$ satisfies, in weak form,
\begin{align}
	\pa{\rho}{t}
	+
	\nabla\cdot(\rho F)
	=
	0
	\label{eq:internal_continuity_equation}
\end{align}
\end{proposition}
\medskip
\noindent
\begin{proof}
For any $f\in C_c^1(\mathcal{M}_{\mathrm{int}})$, integration by parts gives
\begin{align}
	\int_{\mathcal{M}_{\mathrm{int}}}
	\nabla f(\theta)\cdot (\rho F)\,\pd \theta
	=
	-
	\int_{\mathcal{M}_{\mathrm{int}}}
	f(\theta)\,\nabla\cdot(\rho F)\,\pd \theta
	\label{eq:integration_by_parts_internal}
\end{align}
Thus,
\begin{align}
	\int_{\mathcal{M}_{\mathrm{int}}}
	f(\theta)\left(
	\pa{\rho}{t}
	+
	\nabla\cdot(\rho F)
	\right)\pd \theta
	=
	0
	\label{eq:weak_internal_continuity_equation}
\end{align}
holds for arbitrary $f$, which proves the claim.
\end{proof}
\medskip
\noindent
This equation is a Liouville-type equation on the internal state space and shows that the pushforward of the measure under the semiflow $\phi_t$ is expressed at the density level as a conservation-type partial differential equation.
This is also the internal-evolution counterpart of Eq.~(\ref{eq:continuity_equation}) in Section~\ref{sec:measure_theoretic_formulation}.

\subsection{Concrete realization of internal evolution and observation maps}
\label{subsec:internal_evolution_and_observation_maps}

The vector field $F(\theta,t)$ that governs the time evolution of the internal state is concretely given by chemical evolution equations.
In this subsection, we first define the time evolution of the state variables, then introduce dependent variables and observables such as spectra as measurable mappings.
Finally, we express them in a unified manner as pushforwards of measures.

\subsubsection{Chemical evolution equations}
\label{subsubsec:chemical_evolution_equations}

Assume that the internal state includes
\begin{align}
	\bigl(
		M_{\mathrm{gas}},
		M_{\ast},
		Z
	\bigr)
	\label{eq:chemical_state_variables}
\end{align}

\begin{definition}
\label{def:chemical_evolution_system}
Using functions $\psi,E,I,O,y_Z,E_Z$, the system
\begin{align}
	\ta{M_{\mathrm{gas}}}{t}
	&=
	-
	\psi(\theta,t)
	+
	E(\theta,t)
	+
	I(\theta,t)
	-
	O(\theta,t),
	\label{eq:gas_mass_evolution}
	\\
	\ta{M_{\ast}}{t}
	&=
	\psi(\theta,t)-E(\theta,t),
	\label{eq:stellar_mass_evolution}
	\\
	\ta{(Z M_{\mathrm{gas}})}{t}
	&=
	-y_Z(\theta,t)\psi(\theta,t)
	+
	E_Z(\theta,t)
	+
	Z_I I(\theta,t)
	-
	Z O(\theta,t)
	\label{eq:metallicity_evolution}
\end{align}
is called the chemical evolution equations.
\end{definition}
\medskip
\noindent
This system provides a concrete realization of the vector field $F(\theta,t)$ on the internal state space.
That is, the chemical evolution equations provide the physical content of the abstract internal evolution equation (\ref{eq:internal_ode}).
For details on the construction and physical interpretation of each term, see, e.g., Tinsley (1980) \cite{1980FCPh....5..287T}, Matteucci (2012) \cite{Matteucci2012ChemicalEvolution}, Takeuchi (2025a,b) \cite{takeuchi2025physics, takeuchi2025applications}.

\subsubsection{Dust as a dependent variable}
\label{subsubsec:dust_as_dependent_variable}

\begin{fact}
\label{fact:dust_as_function_of_internal_state}
The dust mass $D$ can be expressed as a function of the internal state $\theta$:
\begin{align}
	D=D(\theta)
	\label{eq:dust_as_state_function}
\end{align}
\end{fact}
\medskip
\noindent
Thus, dust is not treated as an independent degree of freedom but as a dependent variable determined by the internal state \citep[see, e.g.,][]{2013EP&S...65..213A, 2013MNRAS.432..637A}.
Such dependent quantities can be treated collectively as measurable functions on the internal state space and can also be regarded as special cases of the observation mappings introduced below.

\subsubsection{Spectral map}
\label{subsubsec:spectral_map}

\begin{definition}
\label{def:spectral_map}
For wavelength $\lambda$, define the measurable mapping
\begin{align}
	\mathcal{S}_\lambda:
	\mathcal{M}_{\mathrm{int}}
	\to
	\mathbb{R}_{+}
	\label{eq:spectral_map}
\end{align}
as the spectral map.
\end{definition}
\medskip
\noindent
The luminosity is given by
\begin{align}
	L_\lambda=\mathcal{S}_\lambda(\theta)
	\label{eq:luminosity_from_spectral_map}
\end{align}

\begin{proposition}
\label{prop:spectral_pushforward_measure}
For a measure $\nu_t$, define
\begin{align}
	\mu_{t,\lambda}=(\mathcal{S}_\lambda)_{\#}\nu_t
	\label{eq:spectral_pushforward_measure}
\end{align}
Then for any bounded measurable function $f:\mathbb{R}_{+}\to\mathbb{R}$,
\begin{align}
	\int_{\mathbb{R}_{+}} f(L)\,\mu_{t,\lambda}(\pd L)
	=
	\int_{\mathcal{M}_{\mathrm{int}}}
	f\bigl(\mathcal{S}_\lambda(\theta)\bigr)\,\nu_t(\pd \theta)
	\label{eq:spectral_pushforward_integral}
\end{align}
holds.
\end{proposition}
\medskip
\noindent
\begin{proof}
This follows by applying Proposition~\ref{prop:pushforward_integral_formula} of Section~\ref{sec:pushforward_structure} with $\Phi=\mathcal{S}_\lambda$.
\end{proof}
\medskip
\noindent
Thus, the spectral distribution is defined as the pushforward of the internal state distribution $\nu_t$ to the observable $L_\lambda$.
This differs from the projection from latent state to observation space via a probabilistic kernel introduced in Section~\ref{sec:measure_theoretic_formulation}, as here we consider a deterministic mapping from internal states to a specific observable.
However, both can be understood in a unified manner as operations that transport a measure to another space.

\subsubsection{Representation of measure evolution}
\label{subsubsec:representation_of_measure_evolution}

From the above,
\begin{align}
	\nu_t=(\phi_t)_{\#}\nu_0,
	\qquad
	\mu_{t,\lambda}
	=
	(\mathcal{S}_\lambda\circ\phi_t)_{\#}\nu_0
	\label{eq:internal_and_spectral_evolution}
\end{align}
holds.

\begin{proposition}
\label{prop:composition_rule_pushforward_internal}
\begin{align}
	(\mathcal{S}_\lambda)_{\#}(\phi_t)_{\#}
	=
	(\mathcal{S}_\lambda\circ\phi_t)_{\#}
	\label{eq:composition_rule_pushforward_internal}
\end{align}
holds.
\end{proposition}
\medskip
\noindent
\begin{proof}
This is the composition rule of pushforward.
For any measurable set $A$,
\begin{align}
	\bigl[(\mathcal{S}_\lambda)_{\#}(\phi_t)_{\#}\nu_0\bigr](A)
	&=
	(\phi_t)_{\#}\nu_0\bigl(\mathcal{S}_\lambda^{-1}(A)\bigr)
	\notag\\
	&=
	\nu_0\bigl(\phi_t^{-1}(\mathcal{S}_\lambda^{-1}(A))\bigr)
	\notag\\
	&=
	\nu_0\bigl((\mathcal{S}_\lambda\circ\phi_t)^{-1}(A)\bigr)
	\notag\\
	&=
	\bigl[(\mathcal{S}_\lambda\circ\phi_t)_{\#}\nu_0\bigr](A)
\end{align}
which proves the claim.
\end{proof}
\medskip
\noindent
Therefore, internal evolution and the generation of observables form a closed structure under the composition of measurable mappings.
This can be regarded as a concrete realization, in terms of deterministic flow on the internal state space and observation mappings, of the measure hierarchy
\begin{align}
	\mu_t\longmapsto \Phi_{\#}\mu_t
\end{align}
introduced in Section~\ref{sec:pushforward_structure}.

\medskip
\noindent
Through the above construction, the internal evolution of galaxies is given as a deterministic semiflow $\phi_t$ on the state space, and observables are generated through the measurable mapping $\mathcal{S}_\lambda$ defined on it.
Since the time evolution is expressed as the pushforward of measures, the evolution on the galaxy manifold is described in a unified manner as the composition of mappings
\begin{align}
	\nu_0
	\longmapsto
	(\phi_t)_{\#}\nu_0
	\longmapsto
	(\mathcal{S}_\lambda\circ\phi_t)_{\#}\nu_0
	\label{eq:internal_evolution_hierarchy}
\end{align}
This structure connects internal physical processes and the generation of observables within the same measure-theoretic framework and shows that the dynamics on the galaxy manifold is closed under the action of measurable mappings.
Moreover, while preserving the primary definition of the galaxy manifold given in Section~\ref{sec:measure_theoretic_formulation}, this section is positioned as the stage in which concrete internal dynamics are introduced on top of it.

\section{Rigorous formulation of mergers as jump processes on the galaxy manifold}
\label{sec:jump_process}

In the previous sections, galaxy evolution has been formulated as the time evolution of a measure $\nu_t$ on the state space.
In particular, in Section~\ref{sec:measure_theoretic_formulation}, the general form of time evolution is given by Eq.~(\ref{eq:weak_evolution_equation}) and Eq.~(\ref{eq:generator_decomposition}), and the decomposition into a continuous transport term and a jump term is introduced by Definition~\ref{def:cont_jump_operators}.
Furthermore, in Section~\ref{sec:internal_evolution}, the continuous transport term is concretized as a pushforward generated by a deterministic semiflow, and its density representation is shown to obey a Liouville-type equation.
However, in galaxy evolution there also exist intrinsically discrete processes that cannot be described by such continuous deformations alone.
A representative example is galaxy mergers, which are understood not as the continuous motion of points on the state space, but as a discontinuous transition in which two points are mapped into a single point.

Such a process must be described not at the level of individual trajectories, but as a change of the measure itself.
That is, a jump process is an object that should be formulated as a nonlinear operator acting on measures.
In this section, based on this standpoint, we rigorously construct the merger process as a jump operator defined on a general measurable space.
The formulation here does not depend on any specific physical model, but is given as a general theory closed in measure-theoretic terms.

In the following, we first introduce a coagulation kernel describing two-body interactions, and define the jump operator based on it.
Next, we organize its weak-form representation in correspondence with the general formulas in Section~\ref{sec:measure_theoretic_formulation}, and further give its decomposition into a rate and a conditional distribution.
We then clarify the structure of additive conserved quantities, and finally position the present formulation through its correspondence with the classical coagulation equation.

\subsection{Construction of the coagulation kernel and the jump operator}
\label{subsec:coagulation_kernel_and_jump_operator}

\subsubsection{Definition of the coagulation kernel}
\label{subsubsec:coagulation_kernel_definition}

Fix a measurable space $(\mathcal{M},\mathcal{F})$.
The state of a galaxy is represented as an element of $\mathcal{M}$, and the merger process is described as a transition from two states $(z_1,z_2)$ to a new state $z$.
To quantify this transition, we introduce a nonnegative measurable function on $\mathcal{M}\times\mathcal{M}\times\mathcal{M}$,
\begin{align}
	K(z_1,z_2\to z;t)\ge0
	\label{eq:coagulation_kernel_section5}
\end{align}
This is the coagulation kernel introduced in Section~\ref{sec:measure_theoretic_formulation}.

\begin{definition}
\label{def:coagulation_kernel_section5}
A function $K$ is called a coagulation kernel if, for any probability measure $\nu\in\mathcal{P}(\mathcal{M})$,
\begin{align}
	\int_{\mathcal{M}}\int_{\mathcal{M}}
	\left(
		\int_{\mathcal{M}}K(z_1,z_2\to z;t)\,\pd z
	\right)
	\nu(\pd z_1)\nu(\pd z_2)
	<
	\infty
	\label{eq:coagulation_kernel_integrability_section5}
\end{align}
holds.
\end{definition}
\medskip
\noindent
This condition guarantees that the total amount of states generated per unit time is finite, and is the basic assumption for the jump operator defined later to be well defined as a finite measure.
Moreover, reflecting the fact that a two-body interaction does not depend on the order of the pair, we assume the symmetry
\begin{align}
	K(z_1,z_2\to z;t)=K(z_2,z_1\to z;t)
	\label{eq:coagulation_kernel_symmetry_section5}
\end{align}
This is the same assumption as Eq.~(\ref{eq:coagulation_kernel_symmetry}) in Section~\ref{sec:measure_theoretic_formulation}, but in the present section we use it to refine the structure of the operator.

Furthermore, following the definition in Section~\ref{sec:measure_theoretic_formulation}, we define the destruction rate by
\begin{align}
	\kappa(z,z';t)
	=
	\int_{\mathcal{M}}K(z,z'\to \tilde{z};t)\,\pd \tilde{z}
	\label{eq:destruction_rate_section5}
\end{align}

\subsubsection{Definition of the jump operator}
\label{subsubsec:jump_operator_definition}

Once the coagulation kernel is given, we define the jump operator acting on measures as follows.

\begin{definition}
\label{def:jump_operator_measure_form}
For a probability measure $\nu\in\mathcal{P}(\mathcal{M})$, define the jump operator $\mathcal{L}_{\mathrm{jump}}[\nu]$ by
\begin{align}
	\mathcal{L}_{\mathrm{jump}}[\nu](\pd z)
	=
	\frac{1}{2}
	\int_{\mathcal{M}}\int_{\mathcal{M}}
	K(z_1,z_2\to z;t)\,\nu(\pd z_1)\nu(\pd z_2)
	-
	\nu(\pd z)
	\int_{\mathcal{M}}\kappa(z,z';t)\,\nu(\pd z')
	\label{eq:jump_operator_measure_form}
\end{align}
\end{definition}
\medskip
\noindent
The first term represents the contribution by which a new state $z$ is generated, while the second term represents the contribution by which the state $z$ is lost through interaction with another state.
The coefficient $1/2$ is introduced in order to correct for the double counting of unordered pairs.
In this way, the jump operator is naturally constructed as the difference between generation and loss.
In Section~\ref{sec:measure_theoretic_formulation}, the jump operator was introduced from the weak form, whereas in the present section we proceed in the reverse order: we first define it directly as a measure-valued operator and then derive its weak form.

\begin{proposition}
\label{prop:jump_operator_finite_signed_measure}
$\mathcal{L}_{\mathrm{jump}}[\nu]$ defines a finite signed measure.
\end{proposition}
\medskip
\noindent
\begin{proof}
Both the generation term and the loss term define nonnegative measures.
By assumption (\ref{eq:coagulation_kernel_integrability_section5}), the total mass of both terms is finite, and therefore their difference, which defines $\mathcal{L}_{\mathrm{jump}}[\nu]$, is a finite signed measure.
\end{proof}
\medskip
\noindent
Therefore, the jump operator is a well-defined nonlinear operator on the space of measures and can be incorporated into the time evolution equation of the measure.
This guarantees that the construction of the present section is not merely a formal description, but is rigorously closed as a measure-theoretic object.

\subsection{Weak form and structural decomposition}
\label{subsec:weak_form_and_structural_decomposition}

\subsubsection{Weak-form representation}
\label{subsubsec:weak_form_representation}

The jump operator can be expressed as a dual action on functions.
This is consistent with the weak-form framework used in Section~\ref{sec:measure_theoretic_formulation}, and is also useful in comparison with the continuous transport term.

\begin{proposition}
\label{prop:jump_operator_weak_form_section5}
For any bounded measurable function $f$,
\begin{align}
	&\int_{\mathcal{M}} f(z)\,\mathcal{L}_{\mathrm{jump}}[\nu](\pd z)\notag \\
	&\qquad =
	\frac{1}{2}
	\int_{\mathcal{M}}\int_{\mathcal{M}}
	\left[
	\int_{\mathcal{M}} f(z)K(z_1,z_2\to z;t)\,\pd z
	-
	f(z_1)
	-
	f(z_2)
	\right]
	\nu(\pd z_1)\nu(\pd z_2)
	\label{eq:jump_operator_weak_form_section5}
\end{align}
holds.
\end{proposition}
\medskip
\noindent
\begin{proof}
Multiplying Eq.~(\ref{eq:jump_operator_measure_form}) of Definition~\ref{def:jump_operator_measure_form} by $f$ and integrating, from the generation term we obtain
\begin{align}
	\frac{1}{2}
	\int_{\mathcal{M}}\int_{\mathcal{M}}
	\left(
	\int_{\mathcal{M}} f(z)K(z_1,z_2\to z;t)\,\pd z
	\right)
	\nu(\pd z_1)\nu(\pd z_2)
\end{align}
On the other hand, using Eq.~(\ref{eq:destruction_rate_section5}), the loss term can be symmetrized as
\begin{align}
	\int_{\mathcal{M}} f(z)\nu(\pd z)
	\int_{\mathcal{M}}\kappa(z,z';t)\,\nu(\pd z')
	=
	\frac{1}{2}
	\int_{\mathcal{M}}\int_{\mathcal{M}}
	\bigl[f(z_1)+f(z_2)\bigr]
	\nu(\pd z_1)\nu(\pd z_2)
\end{align}
Combining the two yields the claim.
\end{proof}
\medskip
\noindent
This equation is nothing other than a rederivation of Eq.~(\ref{eq:jump_operator}) and Eq.~(\ref{eq:jump_operator2}) of Section~\ref{sec:measure_theoretic_formulation} from the measure-valued definition given in the present section.
In particular, since the integrand is evaluated against $\nu\otimes\nu$, it is clear that this operator is intrinsically quadratic.
Moreover, this weak form combines the generation term and the loss term into a single expression, and provides the basic form for discussing conserved quantities and limiting operations.

\subsubsection{Decomposition into rate and conditional distribution}
\label{subsubsec:rate_and_conditional_distribution}

The coagulation kernel can be decomposed into its total mass and a normalized distribution.
This is the most basic decomposition for understanding the structure of the jump process.

\begin{definition}
\label{def:rate_and_conditional_distribution}
\begin{align}
	R(z_1,z_2;t)
	=
	\int_{\mathcal{M}}K(z_1,z_2\to z;t)\,\pd z
	\label{eq:coagulation_rate}
\end{align}
is called the coagulation rate\footnote{
In this paper, we call $R(z_1,z_2;t)$ the ``coagulation rate,'' but this is the quantity corresponding to the jump intensity in the theory of stochastic processes.
That is, $R(z_1,z_2;t)$ represents, for the state pair $(z_1,z_2)$, the total intensity per unit time with which a transition to some state $z$ occurs, and is defined as the total mass of the kernel $K(z_1,z_2\to z;t)$.
In this section, we use the term ``rate'' because it corresponds to physical intuition, but mathematically it should be understood as an intensity.
} and
\begin{align}
	P(z\mid z_1,z_2;t)
	=
	\frac{K(z_1,z_2\to z;t)}{R(z_1,z_2;t)}
	\label{eq:conditional_distribution_jump}
\end{align}
is called the conditional distribution.
However, this is defined on the region where $R(z_1,z_2;t)>0$.
\end{definition}
\medskip
\noindent
\begin{proposition}
\label{prop:kernel_factorization}
When $R(z_1,z_2;t)>0$,
\begin{align}
	K(z_1,z_2\to z;t)
	=
	R(z_1,z_2;t)\,P(z\mid z_1,z_2;t)
	\label{eq:kernel_factorization}
\end{align}
holds, and furthermore
\begin{align}
	\int_{\mathcal{M}}P(z\mid z_1,z_2;t)\,\pd z
	=
	1
	\label{eq:conditional_distribution_normalization}
\end{align}
\end{proposition}
\medskip
\noindent
\begin{proof}
This follows immediately from the definitions.
\end{proof}
\medskip
\noindent
This decomposition separates the jump process into the rate $R$, which gives ``how frequently an interaction occurs,'' and the probability distribution $P$, which gives ``to which state the transition occurs.''
This structure plays an important role later in the analysis of conserved quantities.
Moreover, in the context of the galaxy manifold, this separation corresponds to conceptually distinguishing the merger frequency from the state distribution of the product.

\subsection{Conservation structure and classical limit}
\label{subsec:conservation_and_classical_limit}

\subsubsection{Additive conserved quantities}
\label{subsubsec:additive_conserved_quantities}

Quantities conserved under the jump process can be described as averages with respect to the conditional distribution $P$.
This means that, although the jump operator does not in general conserve particle number, it may possess invariance with respect to appropriate additive quantities.
In Section~\ref{sec:measure_theoretic_formulation}, the conservation of the corresponding moment for an additive conserved quantity $q$ was stated as a proposition, but in the present section we rewrite its condition using the decomposition into the rate and the conditional distribution.

\begin{definition}
\label{def:additive_conserved_quantity_section5}
A measurable function $q$ is called an additive conserved quantity if
\begin{align}
	\int_{\mathcal{M}} q(z)P(z\mid z_1,z_2;t)\,\pd z
	=
	q(z_1)+q(z_2)
	\label{eq:additive_conservation_condition_section5}
\end{align}
holds.
\end{definition}

\begin{proposition}
\label{prop:additive_quantity_conserved_under_jump}
\begin{align}
	\int_{\mathcal{M}} q(z)\,\mathcal{L}_{\mathrm{jump}}[\nu](\pd z)=0
	\label{eq:conservation_of_additive_quantity_under_jump}
\end{align}
holds.
\end{proposition}
\medskip
\noindent
\begin{proof}
Substituting $f=q$ into the weak form (\ref{eq:jump_operator_weak_form_section5}) and using Eq.~(\ref{eq:kernel_factorization}), the integrand becomes
\begin{align}
	R(z_1,z_2;t)
	\left[
	\int_{\mathcal{M}} q(z)P(z\mid z_1,z_2;t)\,\pd z
	-
	q(z_1)-q(z_2)
	\right]
\end{align}
This vanishes by the condition of Definition~\ref{def:additive_conserved_quantity_section5}, and the claim follows.
\end{proof}
\medskip
\noindent
Therefore, quantities satisfying this condition are conserved under the jump process.
This result shows that the jump operator is not merely a description of particle-number change, but a framework that precisely distinguishes which quantities remain invariant.

\begin{fact}
\label{fact:number_not_conserved}
The constant function $q\equiv 1$ is not, in general, an additive conserved quantity.
\end{fact}
\medskip
\noindent
Therefore, particle number is not conserved in general.
This is the measure-theoretic expression of the fact that a merger is a transition from two bodies to one.

\subsubsection{Correspondence with the classical coagulation equation}
\label{subsubsec:classical_coagulation_limit}

The formulation of this section reproduces the classical coagulation equation under an appropriate specialization.
This is important in showing that the present framework is a natural extension that contains the existing theory.

\begin{proposition}
\label{prop:smoluchowski_limit}
\begin{align}
	K(z_1,z_2\to z;t)
	=
	a(z_1,z_2;t)\,\deltadir(z-z_1-z_2)
	\label{eq:smoluchowski_kernel_embedding}
\end{align}
then
\begin{align}
	\int_{\mathcal{M}} f(z)\,\mathcal{L}_{\mathrm{jump}}[\nu](\pd z)
	=
	\frac{1}{2}
	\int_{\mathcal{M}}\int_{\mathcal{M}}
	\left[
	f(z_1+z_2)-f(z_1)-f(z_2)
	\right]
	a(z_1,z_2;t)\nu(\pd z_1)\nu(\pd z_2)
	\label{eq:smoluchowski_weak_form}
\end{align}
results.
\end{proposition}
\medskip
\noindent
\begin{proof}
Substituting Eq.~(\ref{eq:smoluchowski_kernel_embedding}) into the weak form (\ref{eq:jump_operator_weak_form_section5}) and using the delta-function evaluation
\begin{align}
	\int_{\mathcal{M}} f(z)\,\deltadir(z-z_1-z_2)\,\pd z
	=
	f(z_1+z_2)
	\label{eq:delta_evaluation_smoluchowski}
\end{align}
the result follows immediately.
\end{proof}
\medskip
\noindent
This coincides with the weak form of the Smoluchowski-type coagulation equation and shows that the present formulation is its natural generalization.
Accordingly, the coagulation kernel introduced in this section plays the role of connecting general jump processes on the galaxy manifold with classical coagulation theory.
In other words, while containing the known one-dimensional mass coagulation theory as a special limit, the theory of this section provides a structure applicable to higher dimensions and general state spaces.

\subsection{Unified understanding as an operator}
\label{subsec:unified_operator_view}

From the above, the merger process is formulated as the nonlinear operator on the space of measures
\begin{align}
	\nu\longmapsto\mathcal{L}_{\mathrm{jump}}[\nu]
	\label{eq:jump_operator_map}
\end{align}
This operator has a quadratic structure depending on $\nu\otimes\nu$, and has properties that are essentially different from the continuous transport term concretized in Section~\ref{sec:internal_evolution}.
That is, whereas continuous transport is a first-order pushforward generated by a semiflow, the jump operator has a nonlinear structure of generation and loss arising from two-body interactions.
This difference measure-theoretically reflects the essential distinction between the two types of time evolution contained in galaxy evolution, namely continuous deformations and discrete transitions.

Accordingly, the measure evolution on the galaxy manifold is described in a unified way by the evolution equation
\begin{align}
	\ta{\nu_t}{t}
	=
	\mathcal{L}_{\mathrm{cont}}[\nu_t]
	+
	\mathcal{L}_{\mathrm{jump}}[\nu_t]
	\label{eq:full_measure_evolution_with_jump}
\end{align}
Here the former corresponds to the deterministic semiflow concretized in Section~\ref{sec:internal_evolution}, while the latter represents the nonlinear jumps arising from the two-body interactions defined in this section.
As the sum of these two operators, continuous evolution and discrete transitions are integrated within the same measure-theoretic framework.

What is further important is that this unification is not merely a formal composition, but that both are defined as operators on the space of measures.
Accordingly, the time evolution of the galaxy manifold is understood not as a collection of equations of motion for points, but as an operator equation acting on measures.
The continuous transport term describes the smooth deformation of the entire population through the pushforward by a measurable semiflow, while the jump term represents the rearrangement of the measure and mass transport through two-body interactions.
In this sense, the construction of this section is an operator-theoretic extension for adding discrete interactions to the measure evolution on the galaxy manifold, and essentially generalizes the deterministic theory given in the previous sections.

\section{Coalescence process II: many-body coalescence and effective two-body theory}
\label{sec:many_body_coalescence}

\begin{figure}[t]
\centering
\includegraphics[width=\textwidth]{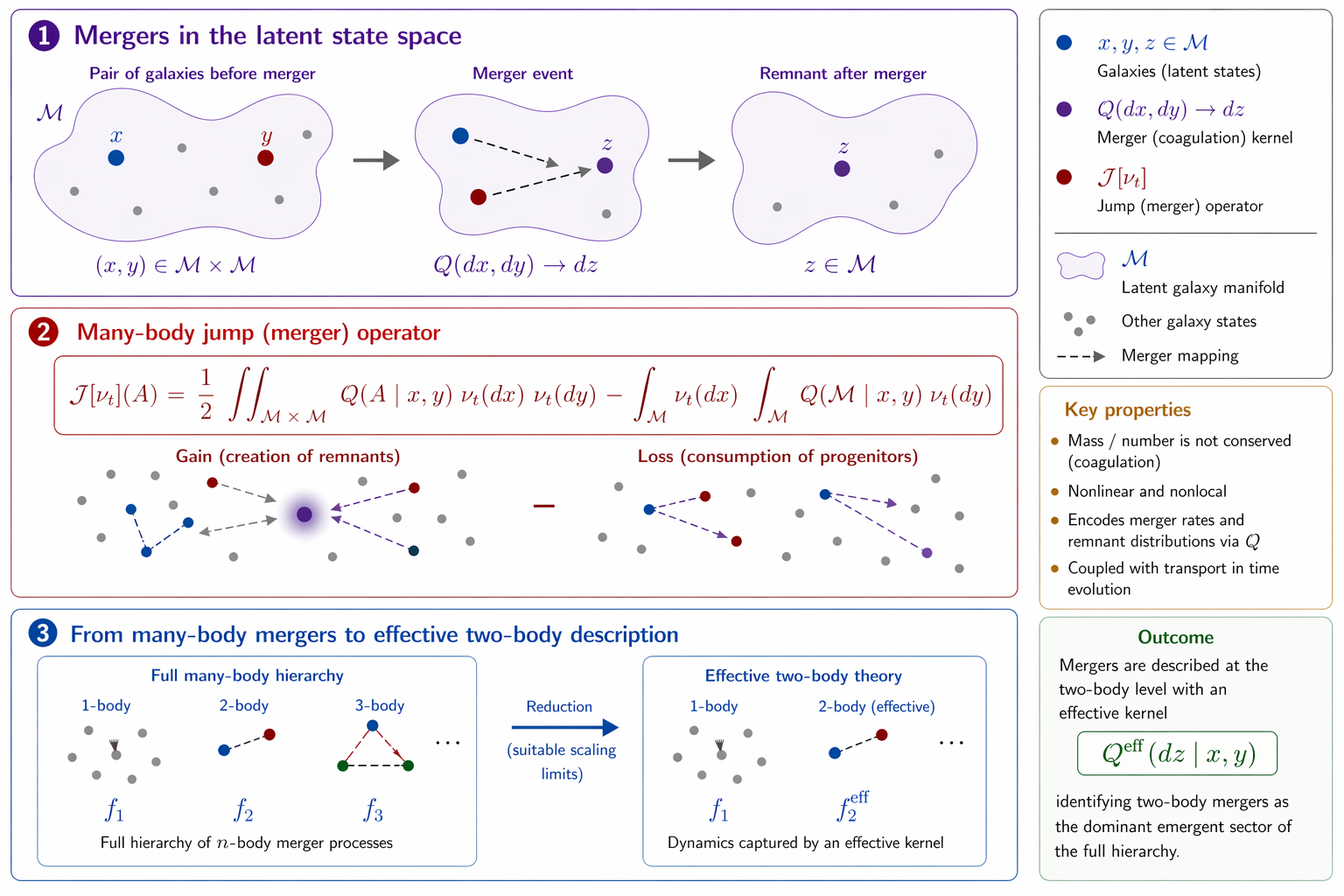}
\caption{
Schematic illustration of merger processes in the latent galaxy state space and their effective description. 
Top: a merger event is represented as a transition from a pair of progenitor states $(x,y)\in\mathcal{M}\times\mathcal{M}$ to a remnant state $z\in\mathcal{M}$ through the coagulation kernel $\mathcal{Q}(\pd z\mid x,y)$. 
Middle: the corresponding jump operator $\mathcal{J}[\nu_t]$ consists of a gain term describing the creation of remnants and a loss term describing the consumption of progenitors, yielding a nonlinear and nonlocal contribution to the time evolution of the population measure. 
Bottom: the full many-body interaction hierarchy can, under suitable scaling limits, be reduced to an effective two-body theory with an effective kernel $\mathcal{Q}^{\mathrm{eff}}$, which captures the dominant large-scale merger dynamics.
}
\label{fig:merger_hierarchy}
\end{figure}

In this section, we extend the two-body coalescence process formulated in Section~\ref{sec:jump_process} and provide a general theory that allows many-body coalescence from $n$ bodies to one body.
The theory of the previous section is positioned as the lowest-order approximation within the framework of many-body interactions constructed in this section.
Accordingly, the purpose of this section is not merely to provide an extension, but to clarify that jump processes on the galaxy manifold are essentially understood as a hierarchical structure of many-body interactions.
Moreover, since physically higher-order many-body coalescences are expected to be probabilistically suppressed, the many-body theory is reduced to an effective two-body theory in an appropriate limit.
In this sense, this section also plays the role of providing the correspondence between the general theory and practical approximation.
An additional important point is that, by extending to many-body interactions, the jump operator is extended from a merely quadratic nonlinearity to a series of nonlinear operators of arbitrary order, thereby making explicit that the structure of the dynamics on the space of measures is essentially hierarchical.
This hierarchical structure of merger interactions and its effective reduction are illustrated in Fig.~\ref{fig:merger_hierarchy}.

\subsection{Many-body coalescence kernel}
\label{subsec:many_body_kernel}

Fix $n\ge2$, and let
\begin{align}
	K_n(z_1,\dots,z_n\to z;t)\ge0
	\label{eq:many_body_kernel}
\end{align}
be a nonnegative measurable function on $\mathcal{M}^{n}\times\mathcal{M}$, called the $n$-body coalescence kernel.
This represents the intensity density per unit time with which the $n$ states $(z_1,\dots,z_n)$ transition to the state $z$.
What is important here is that $K_n$ is not merely a function, but is defined as an interaction intensity that has measure-theoretic meaning by being integrable with respect to the measure $\nu^{\otimes n}$ on $\mathcal{M}^n$.
Since the ordering of particles is not essential, we assume that for any permutation $\sigma\in S_n$,
\begin{align}
	K_n(z_{\sigma(1)},\dots,z_{\sigma(n)}\to z;t)
	=
	K_n(z_1,\dots,z_n\to z;t)
	\label{eq:many_body_symmetry}
\end{align}
Furthermore, for any $\nu\in\mathcal{P}(\mathcal{M})$, we assume
\begin{align}
	\int_{\mathcal{M}^{n}}
	\left(
		\int_{\mathcal{M}}K_n(z_1,\dots,z_n\to z;t)\,\pd z
	\right)
	\nu^{\otimes n}(\pd z_1\cdots \pd z_n)
	<
	\infty
	\label{eq:many_body_integrability}
\end{align}
This condition is the necessary and sufficient condition for the generation term of the jump operator to be defined as a finite measure, and plays the role of controlling the nontrivial problem of integrability that arises upon extending to many-body interactions.

\subsection{$n$-body jump operator}
\label{subsec:n_body_jump_operator}

The jump operator corresponding to $n$-body coalescence is defined in weak form by
\begin{align}
	&\int_{\mathcal{M}} f(z)\,\mathcal{L}_{\mathrm{jump}}^{(n)}[\nu](\pd z) \notag\\
	&\qquad =
	\frac{1}{n!}
	\int_{\mathcal{M}^{n}}
	\left[
		\int_{\mathcal{M}} f(z)K_n(z_1,\dots,z_n\to z;t)\,\pd z
		-
		\sum_{i=1}^{n}f(z_i)
	\right]
	\nu^{\otimes n}(\pd z_1\cdots \pd z_n)
	\label{eq:n_body_jump_operator}
\end{align}
In this definition, the first term corresponds to the generation term, and the second term corresponds to the destruction term, so that the change in the measure due to the jump is given as their difference.
The coefficient $1/n!$ is the correction for overcounting based on symmetry, and by this the operator is defined in a natural form with respect to $\nu^{\otimes n}$.
What is important here is that $\mathcal{L}_{\mathrm{jump}}^{(n)}$ is an $n$th-order nonlinear operator with respect to $\nu$, and this is the direct generalization of the quadratic nonlinearity in the two-body case, but mathematically it means that the order of the nonlinear operator is extended to infinity.

At this point, the full jump operator is written as
\begin{align}
	\mathcal{L}_{\mathrm{jump}}[\nu]
	=
	\sum_{n=2}^{\infty}
	\mathcal{L}_{\mathrm{jump}}^{(n)}[\nu]
	\label{eq:many_body_jump_expansion}
\end{align}
Although this series representation is formally simple, strictly speaking it involves issues of convergence and domain of definition.
That is, even if each term is finite, in order for the operator as an infinite sum to be well defined, convergence is required in an appropriate norm or in a weak topology.
In this paper, we avoid this problem by assuming a situation in which the higher-order terms are physically suppressed, but this point is an essential mathematical issue in the rigorous analysis of many-body jump processes.
Accordingly, the jump process is understood simultaneously as a hierarchical expansion of many-body interactions and as a series structure of nonlinear operators.

\subsection{Generalization of conserved quantities}
\label{subsec:many_body_conservation}

When a measurable function $q$ satisfies
\begin{align}
	\int_{\mathcal{M}}q(z)\,P_n(z\mid z_1,\dots,z_n;t)\,\pd z
	=
	\sum_{i=1}^{n}q(z_i)
	\label{eq:many_body_conservation}
\end{align}
$q$ is an additive conserved quantity for $n$-body coalescence.
This condition means that $q$, as the expectation value of the generated state, agrees with the sum over the initial states, and physically corresponds to mass conservation or other conservation laws.
In this case,
\begin{align}
	\int_{\mathcal{M}}q(z)\,\mathcal{L}_{\mathrm{jump}}^{(n)}[\nu](\pd z)=0
\end{align}
holds.
Accordingly, even when many-body interactions are allowed, the structure of conserved quantities is maintained in the same form as in the two-body case.
This invariance shows that the jump operator is not merely a nonlinear transformation, but is a constrained operator in the sense that it leaves specific linear functionals invariant.

\subsection{Many-body expansion, convergence, effective theory, and response coefficients}
\label{subsec:many_body_effective_response}

Although the importance of many-body interactions is generally expected to decrease as $n$ increases, in order to position this intuition clearly within the theory, it is necessary to consider simultaneously the convergence of the jump operator as a series expansion and the structure of its dominant terms. Accordingly, in this section, we discuss in a unified way the scaling of the many-body kernels, the convergence conditions for the jump operator as a series, the reduction to an effective two-body theory, and the associated structure of response coefficients.

First, introduce a small parameter $\varepsilon$ and assume
\begin{align}
	K_n = O(\varepsilon^{\,n-2})K_n^{(0)}
	\label{eq:many_body_scaling_refined}
\end{align}
This scaling is a natural assumption based on local occupation number or collision probability, and $\varepsilon$ is interpreted as a dimensionless quantity characterizing the diluteness of the system. What is important here is that this assumption is not merely a formal expansion, but has the role of guaranteeing the convergence of the jump operator as a series. Indeed, considering the weak form for any bounded measurable function $f$,
\begin{align}
	\left|
	\int f(z)\,\mathcal{L}_{\mathrm{jump}}^{(n)}[\nu](\pd z)
	\right|
	\le
	C\,\varepsilon^{\,n-2}\|f\|_{\infty}
\end{align}
is expected to hold, and therefore under $\varepsilon<1$,
\begin{align}
	\sum_{n=2}^{\infty}
	\mathcal{L}_{\mathrm{jump}}^{(n)}[\nu]
\end{align}
converges at least in the weak topology.
More concretely, if $\sup_n \varepsilon^{\,n-2}\|K_n^{(0)}\|$ is controllable, the jump operator can be defined as an operator series on a Banach space.
In this sense, the many-body expansion is not merely a formal series, but provides a well-defined dynamics under appropriate decay conditions.

At this point, since the jump operator is expanded as
\begin{align}
	\mathcal{L}_{\mathrm{jump}}
	=
	\mathcal{L}_{\mathrm{jump}}^{(2)}
	+
	O(\varepsilon)\mathcal{L}_{\mathrm{jump}}^{(3)}
	+
	O(\varepsilon^2)\mathcal{L}_{\mathrm{jump}}^{(4)}
	+\cdots
\end{align}
in the limit $\varepsilon\ll1$,
\begin{align}
	\mathcal{L}_{\mathrm{jump}}[\nu]
	\approx
	\mathcal{L}_{\mathrm{jump}}^{(2)}[\nu]
\end{align}
holds.
This limit corresponds to the low-density or weak-interaction limit, in which many-body effects are organized as higher-order corrections.
Accordingly, two-body coalescence appears naturally not as a special assumption within the many-body theory, but as the lowest-order term of a convergent many-body expansion.

Next, consider the problem of reducing many-body interactions to an effective two-body theory.
In general, since the terms with $n\ge3$ depend on $\nu^{\otimes n}$, they do not in themselves form a closed two-body theory.
However, by statistically averaging or coarse-graining these higher-order terms, they can be absorbed as a renormalization into the two-body kernel.
That is, introducing
\begin{align}
	K_{\mathrm{eff}}(z_1,z_2\to z;t)
	=
	K_2(z_1,z_2\to z;t)
	+
	\Delta K_2(z_1,z_2\to z;t)
\end{align}
the contributions of three-body and higher interactions are included in $\Delta K_2$.
At this point, since $\Delta K_2$ generally depends on $\nu$ and on its higher-order correlation functions, the reduction is strictly speaking not closed, but under weak correlation or mean-field approximation, an effective two-body theory is obtained.
This structure corresponds to the typical renormalization structure from microscopic many-body interactions to macroscopic effective theory.

Furthermore, the present structure is also naturally understood from the viewpoint of response coefficients.
The response coefficient introduced in Section~\ref{sec:measure_theoretic_formulation},
\begin{align}
	R = \frac{\partial \ln N}{\partial \ln \beta}
\end{align}
was the quantity defined as the response of an observable to a deformation of the measure, but under the many-body expansion of the jump operator, this response has a hierarchical structure.
That is, since $\mathcal{L}_{\mathrm{jump}}^{(n)}$ depends on $\nu^{\otimes n}$, the response coefficient is also decomposed as a sum of many-body contributions of the form
\begin{align}
	R = R^{(2)} + R^{(3)} + R^{(4)} + \cdots
\end{align}
Here $R^{(n)}$ is the response component originating from the $n$-body interaction, and under the scaling assumption (\ref{eq:many_body_scaling_refined}),
\begin{align}
	R^{(n)} = O(\varepsilon^{\,n-2})
\end{align}
is obtained.
Accordingly, in the low-density limit,
\begin{align}
	R \approx R^{(2)}
\end{align}
holds, and the response coefficient is likewise dominated by the two-body process.
This result shows that the response of observables directly reflects the hierarchical structure of the dynamics.

Summarizing the above, the many-body jump process has the structure
\begin{align}
\begin{array}{c}
\text{many-body operator series}
\\[3pt]
\big\downarrow
\\[3pt]
\text{control by convergence conditions}
\\[3pt]
\big\downarrow
\\[3pt]
\text{reduction to an effective two-body theory}
\\[3pt]
\big\downarrow
\\[3pt]
\text{hierarchical decomposition of response coefficients}
\end{array}
\end{align}
Through this framework, both in dynamics and in observables, many-body effects and their effective description are treated in a unified way.
What is particularly important is that the two-body theory appears not as a mere approximation, but as the dominant term of a convergent many-body expansion, and thereby the basic formulation of this paper is justified both theoretically and physically.

\subsection{Concept of solution to the measure evolution equation}
\label{subsec:solution_concept}

In the previous sections, galaxy evolution has been formulated as the evolution equation on the space of measures
\begin{align}
	\ta{\nu_t}{t}
	=
	\mathcal{L}[\nu_t],
	\qquad
	\mathcal{L}
	=
	\mathcal{L}_{\mathrm{cont}}
	+
	\mathcal{L}_{\mathrm{jump}} \notag
\end{align}
Here $\mathcal{L}_{\mathrm{cont}}$ is the continuous transport operator corresponding to the deterministic semiflow, and $\mathcal{L}_{\mathrm{jump}}$ is the nonlinear jump operator originating from many-body interactions (see Section~\ref{sec:measure_theoretic_formulation}, Eq.~\eqref{eq:jump_operator}).
For this equation, it is natural to define the solution in weak form with respect to measures.

\begin{definition}
Given an initial measure $\nu_0\in\mathcal{P}(\mathcal{M})$, a measure-valued mapping $t\mapsto\nu_t\in\mathcal{P}(\mathcal{M})$ is called a weak solution if, for any bounded measurable function $f:\mathcal{M}\to\mathbb{R}$,
\begin{align}
	\ta{}{t}
	\int_{\mathcal{M}} f(z)\,\nu_t(\pd z)
	=
	\int_{\mathcal{M}} f(z)\,\mathcal{L}_{\mathrm{cont}}[\nu_t](\pd z)
	+
	\int_{\mathcal{M}} f(z)\,\mathcal{L}_{\mathrm{jump}}[\nu_t](\pd z)
\end{align}
holds, and furthermore $\nu_{t=0}=\nu_0$ is satisfied.
\end{definition}

Furthermore, in this work we assume at least the weak continuity
\begin{align}
	t\longmapsto
	\int_{\mathcal{M}} f(z)\,\nu_t(\pd z)
\end{align}
for any $f\in C_b(\mathcal{M})$.
By this, $\nu_t$ is defined as a curve on the measure space $\mathcal{P}(\mathcal{M})$.

Corresponding to the operator decomposition, the solution is understood as a superposition of continuous transport and jump process.
That is, the continuous part is given by the semiflow $\phi_t$ as
\begin{align}
	\nu_t^{(\mathrm{cont})}
	=
	(\phi_t)_{\#}\nu_0
\end{align}
and the jump term deforms the measure nonlinearly through many-body interactions.
Accordingly, the general solution $\nu_t$ is characterized as a measure evolution in which deterministic transport and probabilistic interactions act simultaneously.

In addition, for an additive conserved quantity $q$,
\begin{align}
	\ta{}{t}
	\int_{\mathcal{M}} q(z)\,\nu_t(\pd z)
	=
	\int_{\mathcal{M}} q(z)\,\mathcal{L}_{\mathrm{cont}}[\nu_t](\pd z)
\end{align}
holds, and in particular if $\mathcal{L}_{\mathrm{cont}}$ conserves $q$, then this quantity is time-invariant.

From the above, the solution in the present theory is not the trajectory of each individual galaxy, but the time evolution of the measure $\nu_t$ representing the state distribution, and its dynamics is given as a nonlinear operator equation integrating continuous transport and many-body jump processes.
Through this viewpoint, the galaxy manifold is understood not as a static geometric object, but as a dynamical system on the space of measures.

\section{Galaxy evolution as transport under the curvature--dimension condition CD$(K,\infty)$}
\label{sec:cd_transport}

In this section, we place the measure evolution equation of galaxy evolution constructed in the previous sections within the framework of metric measure spaces satisfying the curvature--dimension condition CD$(K,\infty)$, and clarify the geometric and variational structures that hold on that basis (see Appendix~\ref{app:cd} for details).
The focus of this section is to extract the constraints and consequences that appear only when measure evolution is understood as transport in Wasserstein geometry, in particular the contractivity and stability arising from the geodesic convexity of H-function (defined as a negative of entropy \citep[e.g.,][]{villani2009optimal}).
From this viewpoint, galaxy evolution is characterized not as a mere dynamical equation, but as a \emph{reaction--transport system} in which transport geometry, energy dissipation, and curvature control are coupled.
This is not merely a formal reorganization, but means that the admissible trajectories of evolution are geometrically constrained by the introduction of the curvature--dimension condition, and it follows that the formulation by Wasserstein geometry imposes essential constraints on the very pathways of galaxy evolution.
The geometric interpretation of the measure evolution discussed in this section is summarized in Fig.~\ref{fig:transport_geometry}.

\begin{figure}[tp]
\centering
\includegraphics[width=\textwidth]{galaxy_manifold_CD.png}
\caption{
Geometric structure of measure evolution in the Wasserstein space. 
Top: probability measures are connected by geodesic transport paths. 
Middle: under the curvature--dimension condition CD$(K,\infty)$, nearby evolutions contract in time. 
The contraction rate is controlled by the lower curvature bound $K$. 
Bottom: this geometric control yields stable trajectories and stable projected observables. 
}
\label{fig:transport_geometry}
\end{figure}

\subsection{Introduction of metric measure spaces and transport geometry}

Introduce a metric $d$ and a reference measure $\mathfrak{m}$ on the state space $\mathcal{M}$, and let $(\mathcal{M},d,\mathfrak{m})$ be a Polish metric space together with a $\sigma$-finite Borel measure on it.
Furthermore, assume that $\nu_t\in\mathcal{P}_2(\mathcal{M})$ and $\nu_t\ll\mathfrak{m}$, and write
\begin{align}
	\nu_t(\pd z)=\rho(z,t)\,\mathfrak{m}(\pd z)
\end{align}

In this case, the quadratic Wasserstein distance on the probability measure space $\mathcal{P}_2(\mathcal{M})$ is defined by
\begin{align}
	W_2^2(\nu_0,\nu_1)
	=
	\inf_{\pi\in\Pi(\nu_0,\nu_1)}
	\int_{\mathcal{M}\times\mathcal{M}}
	d(z,z')^2\,\pi(\pd z,\pd z')
\end{align}
Through this introduction, the measure space becomes not merely a function space, but a geometric object endowed with a metric structure.
Accordingly, galaxy evolution $\{\nu_t\}$ is defined as a curve on the space of measures, and its deformation is quantified with respect to the distance.

Furthermore, under the assumption of absolute continuity, the evolution is described by the continuity equation
\begin{align}
	\pa{\rho}{t}
	+
	\nabla\cdot(\rho v)
	=
	0
\end{align}
Through this relation, measure evolution is understood in a unified manner not as the motion of individual particles, but as the transport of a mass distribution.
In particular, it is important that, within this framework, the evolution is treated as a deformation consistent with the distance structure defined by optimal transport.

\subsection{Gradient-flow structure and variational formulation}\label{subsec:gradient_flow}

In order to understand measure evolution more structurally, introduce a functional $\mathcal{F}:\mathcal{P}_2(\mathcal{M})\to(-\infty,\infty]$. Here $\mathcal{F}$ is assumed to be lower semicontinuous and convex in an appropriate sense. In this case, the evolution is defined by the sequential optimization problem, for time step $\tau>0$,
\begin{align}
	\nu^{k+1}
	\in
	\operatorname*{argmin}_{\nu}
	\left\{
	\frac{1}{2\tau}W_2^2(\nu,\nu^k)
	+
	\mathcal{F}(\nu)
	\right\}
\end{align}
This scheme was introduced by Jordan--Kinderlehrer--Otto \citep{Jordan1998}, and constructs evolution as the competition between optimal transport and energy minimization in discrete time.
In the limit $\tau\to0$, the continuous curve obtained gives the Wasserstein gradient flow of $\mathcal{F}$.

By this variational construction, the continuous transport term is characterized not as a mere deterministic deformation, but as a
\emph{geometric relaxation process along the direction of steepest descent that decreases the free energy $\mathcal{F}$}.
That is,
\begin{align}
	\ta{\nu_t}{t}
	=
	-
	\nabla_{W_2}\mathcal{F}[\nu_t]
\end{align}
is a concise notation expressing this energy-dissipation structure\footnote{It should be noted that the 'energy' referred to here is the free-energy functional $\mathcal{F}$ in the sense of a Lyapunov functional for the gradient flow. 
This evolution represents a dissipative process minimizing the functional $\mathcal{F}$, rather than the conservation of internal physical energy. 
In the context of the galaxy manifold, this dissipation corresponds to the relaxation of the population distribution toward a geometrically stable state under the curvature constraint.}.
Furthermore, under this structure, the energy-dissipation inequality
\begin{align}
	\mathcal{F}(\nu_t)
	+
	\frac{1}{2}\int_0^t |\dot{\nu}_s|^2\,\pd s
	\le
	\mathcal{F}(\nu_0)
\end{align}
holds, showing that the evolution is controlled as a monotone energy-decreasing process.
Here $|\dot{\nu}_t|$ is the metric speed with respect to the Wasserstein distance.

This result shows that continuous transport is not an arbitrary flow, but a variational dynamics generated by a certain functional $\mathcal{F}$, and that its time evolution is simultaneously controlled by energy dissipation and the geometry of optimal transport.
In this sense, the continuous part of galaxy evolution is positioned as a structure satisfying both dynamical description and variational principle.

\subsection{Curvature--dimension condition and geometric control of evolution}\label{subsec:geometric_control}

In the gradient-flow structure introduced in the previous subsection, the specific choice of the functional $\mathcal{F}$ governing the evolution plays an important role. 
In particular, as a functional consistent with the geometry on the space of measures, introduce the relative H-function
\begin{align}
	\mathcal{H}_{\mathfrak{m}}(\nu)
	=
	\int_{\mathcal{M}}\rho\ln\rho\,\mathfrak{m}(\pd z)
\end{align}
\citep{villani2009optimal}.
Here $\rho$ is the density of $\nu$ with respect to $\mathfrak{m}$. 
This functional quantifies the degree of diffusion or mixing of the measure, and has a fundamental role in transport geometry.

In the context of the gradient flow discussed in Section~\ref{subsec:gradient_flow}, the free-energy functional $\mathcal{F}$ is primarily composed of this H-functional, i.e., $\mathcal{F}[\mathcal{\nu}] = H ( \nu) + \mathcal{V}(\nu)$, where $\mathcal{V}$ represents an external potential or confinement term on the manifold. 
In the following discussion, we focus on the case where the evolution is dominated by the H-functional, which characterizes the purely geometric relaxation of the galaxy population.

When the metric measure space $(\mathcal{M},d,\mathfrak{m})$ satisfies the curvature--dimension condition CD$(K,\infty)$, this functional $\mathcal{H}_{\mathfrak{m}}$ is $K$-convex along Wasserstein geodesics.
That is, for any geodesic $\{\nu_t\}$, $\mathcal{H}_{\mathfrak{m}}(\nu_t)$ is controlled quadratically.

This convexity is not merely a geometric property, but is directly reflected in the dynamics of the gradient flow.
In fact, the gradient flow of $\mathcal{H}_{\mathfrak{m}}$ satisfies the contractivity
\begin{align}
	W_2(\nu_t,\tilde{\nu}_t)
	\le
	e^{-Kt}W_2(\nu_0,\tilde{\nu}_0)
\end{align}
This inequality shows that evolutions starting from different initial conditions approach each other exponentially in time, thereby providing a quantitative guarantee of the stability of the evolution.
Accordingly, the continuous part of galaxy evolution is understood not as mere transport, but as a geometric relaxation process controlled by the lower curvature bound $K$.
In particular, when $K>0$, the convergence is strong, while when $K<0$, diffusive behavior becomes dominant.
In this way, curvature appears as a fundamental quantity characterizing the dynamics of the evolution itself.

This type of stability arises only when the CD$(K,\infty)$ condition is imposed, through which measure evolution is positioned as a system controlled under geometric constraints, making possible a quantitative understanding of its long-time behavior.

\subsection{Unified structure as a reaction--transport system and new consequences}

Integrating the above discussion, galaxy evolution is described by
\begin{align}
    \frac{\partial \nu_t}{\partial t}
    =
    - \nabla_{W_2}\mathcal{F}[\nu_t]
    + \mathcal{L}_{\mathrm{jump}}[\nu_t]
\end{align}
This equation is a reaction--transport system given as the \emph{coupling of a gradient flow in Wasserstein geometry and a jump process based on many-body interactions}.

\subsubsection{Hierarchy of interactions and effective closure}

The jump operator has the many-body expansion
\begin{align}
    \mathcal{L}_{\mathrm{jump}}
    =
    \sum_{n\ge2}\mathcal{L}_{\mathrm{jump}}^{(n)}
\end{align}
and each term is defined as an $n$th-order nonlinear operator depending on $\nu^{\otimes n}$.
However, in the low-density or weak-interaction limit,
\begin{align}
    \mathcal{L}_{\mathrm{jump}}[\nu]
    \approx
    \mathcal{L}_{\mathrm{jump}}^{(2)}[\nu]
\end{align}
holds, and the evolution is described as a closed system depending only on the measure $\nu_t$.

What is important here is that this closure appears naturally not as a mere computational approximation, but as a structure consistent with transport geometry.
That is, two-body coalescence theory is positioned not only as the lowest-order term of many-body interactions, but also as an effective theory compatible with geometric structure.

\subsubsection{Response of observables and geometric sensitivity}

Next, we clarify the relation between this dynamical structure and observables.
For the observed measure introduced in Section~\ref{sec:measure_theoretic_formulation},
\begin{align}
    \mu_t(B)
    =
    \int_{\mathcal{M}}K(B\mid z)\,\nu_t(dz)
\end{align}
any observable statistic $N$ is given in the form
\begin{align}
    N(t)=\int_{\mathcal{X}}\psi(x)\,\mu_t(dx)
\end{align}
Here $\psi$ is a measurable function representing a selection function or weight.

At this point, the time evolution of $N(t)$ is induced through the deformation of $\nu_t$.
In particular, considering a deformation of the system by an external parameter $\beta$, the response of the observable is defined by
\begin{align}
    R=\frac{\partial \ln N}{\partial \ln \beta}
\end{align}
This quantity is a dimensionless quantity representing the sensitivity of the observable statistic to the deformation of the measure, and plays the role of mapping the structure of evolution to observables.
Corresponding to the decomposition of the evolution operator, the response coefficient is also decomposed as
\begin{align}
    R = R_{\mathrm{geom}} + R_{\mathrm{int}}
\end{align}
Here $R_{\mathrm{geom}}$ is the contribution originating from the gradient flow, namely Wasserstein transport, and reflects the geometric structure of the space of measures.
On the other hand, $R_{\mathrm{int}}$ originates from the jump operator and represents the effect of rearrangement by many-body interactions.
In particular, under the CD$(K,\infty)$ condition, since the gradient-flow part is controlled by the lower curvature bound, $R_{\mathrm{geom}}$ is understood as a geometrically constrained response.
Accordingly, the response of observables is positioned not as a mere empirical quantity, but as an integrated indicator reflecting both the geometric structure of the space of measures and many-body interactions.

\subsubsection{Physical implications of geometric constraints}

Here we consider some physical implications of the geometric constraints.
To clarify the physical significance of the ${\rm CD}(K, \infty)$ condition, let us consider the total stellar mass $M_*$ as a primary component of the state variable $z$. 
In this framework, the additive conserved quantity $q(z)=M_*$ ensures that the total mass is preserved during discrete merger events, while the $\mathcal{L}_{\rm jump}$ operator induces a rapid redistribution of the population measure toward higher mass scales.

The $K$-convexity of the H-functional then plays a crucial role in the subsequent relaxation phase. 
Physically, the lower curvature bound $K$ characterizes the ``strength'' of the environmental and internal feedback processes that drive the galaxy population toward a stable scaling relation, such as the star-formation main sequence or the mass-metallicity relation.
If $K>0$, the galaxy population is geometrically forced to converge exponentially toward a unique stable distribution. 
In contrast, a negative $K$ would imply that stochastic processes or environmental instabilities dominate, leading to a broader dispersion in the observed scaling relations. 
Thus, galaxy evolution is understood as a cycle of ``jumps'' in the Wasserstein space followed by a geometrically constrained ``relaxation'' toward structural stability.

\subsubsection{Summary of principles for admissible evolution}

From the above, the conclusions obtained in this section can be summarized as follows.
That is, galaxy evolution is described as a reaction--transport system in which a many-body jump operator is added to a Wasserstein gradient flow on the space of measures satisfying the curvature--dimension condition CD$(K,\infty)$.
Through this structure, structure formation is understood as geometric transport, and galaxy formation as many-body interaction, in a unified manner within the same measure evolution equation.
What is especially important is that this framework does not merely describe galaxy evolution, but imposes constraints on the very class of admissible evolutions.
More specifically, the evolution is controlled by the following principles:

\begin{itemize}
\item \textbf{Variational structure:}
Continuous transport is given as the gradient flow of the free energy $\mathcal{F}$, and the evolution is restricted to the direction that decreases the energy\footnote{It should be noted that the `energy' referred to here is the free-energy functional $\mathcal{F}$ in the sense of a Lyapunov functional for the gradient flow. This evolution represents a dissipative process minimizing the functional $\mathcal{F}$, rather than the conservation of internal physical energy. In the context of the galaxy manifold, this dissipation corresponds to the relaxation of the population distribution toward a geometrically stable state under the curvature constraint.}.
Accordingly, unstable circulatory flows or evolutions in which the energy increases are not allowed.

\item \textbf{Curvature constraint:}
By the CD$(K,\infty)$ condition, evolution on the space of measures has contractivity and stability, and differences in the initial conditions are controlled over time.
As a result, unlimited branching or disordered diffusion is geometrically suppressed.

\item \textbf{Effective closure of interactions:}
In the low-density or weak-interaction limit, the many-body jump operator is effectively reduced to the two-body term, and the degrees of freedom of higher-order interactions are suppressed.

\item \textbf{Restriction of trajectories:}
Under the above structure, the evolution is restricted not to arbitrary measure deformations, but to a set of trajectories consistent with geometric and variational principles.
\end{itemize}
Consequently, the present theory is not only a framework describing galaxy evolution, but also provides a principle that selects its possible evolutionary paths.
Accordingly, the formulation obtained in this study characterizes galaxy evolution as a geometrically controlled dynamics, and is positioned as a theory of constrained evolution.
Furthermore, within this framework, the statistical properties and responses of the observed galaxy distribution are understood in a unified manner as quantities controlled by both the geometric structure of the space of measures and the interactions.
That is, the present theory not only describes galaxy evolution, but \emph{provides a principle that selects the class of its admissible evolutions.}

\section{Conclusion and Outlook}
\label{sec:conclusion}

\subsection{Conclusion}

In this paper, we formulated galaxy evolution as a time evolution on the space of measures, and systematically constructed its structure as an integration of measure theory, operator theory, and transport geometry.

First, by defining a galaxy population as a probability measure $\nu_t\in\mathcal{P}(\mathcal{M})$ on the state space and describing observation as a linear projection by a probability kernel, we clearly separated latent states from observed distributions. Within this framework, finite samples and continuous distributions are unified through empirical measures, and it was shown that observational data are naturally embedded as measure-theoretic objects.

Next, the time evolution was given by the generator
\begin{align}
\mathcal{L}
=
\mathcal{L}_{\mathrm{cont}}
+
\mathcal{L}_{\mathrm{jump}}
\end{align}
and the decomposition into continuous transport and jump process was introduced. The continuous transport was concretized as the pushforward by a deterministic semiflow, and it was shown that, under a density representation, it obeys a conservation-type continuity equation. On the other hand, the jump operator was constructed as a nonlinear operator based on two-body interactions, and its weak-form representation made clear that it has a quadratic structure depending on the product measure.

Furthermore, through the decomposition of the coagulation kernel into the rate and the conditional distribution, it was shown that the structure of the jump process is understood as the product of ``frequency'' and ``transition distribution,'' and the invariance of additive conserved quantities was formulated as a constraint on measure evolution. This structure naturally explains the characteristic feature of galaxy evolution that allows simultaneous change in number density and preservation of conserved quantities.

Next, a general theory including many-body coalescence was introduced, and it was shown that the jump operator has the hierarchical expansion
\begin{align}
\mathcal{L}_{\mathrm{jump}}
=
\sum_{n\ge2}\mathcal{L}_{\mathrm{jump}}^{(n)}
\end{align}
Under this structure, it was clarified that when higher-order interactions are suppressed by scaling, the theory is effectively reduced to a two-body theory. Accordingly, two-body coalescence is positioned not as a mere approximation, but as a dominant structure that appears naturally within the many-body theory.

Finally, by introducing a metric structure on the state space and embedding the measure evolution into the framework of Wasserstein geometry, it was shown that galaxy evolution is described as the reaction--transport system
\begin{align}
\ta{\nu_t}{t}
=
-
\nabla_{W_2}\mathcal{F}[\nu_t]
+
\mathcal{L}_{\mathrm{jump}}[\nu_t]
\end{align}
Furthermore, by introducing the curvature--dimension condition CD$(K,\infty)$, it was clarified that the continuous transport part behaves as a geometrically controlled gradient flow, and that the stability, contractivity, and convergence of the evolution are quantitatively constrained by the lower curvature bound $K$.

These results show that galaxy evolution is not an arbitrary dynamics, but a process constrained by the following three structures:
\begin{itemize}
\item closure as a measure evolution (dynamics on the space of probability measures)
\item variational structure by transport geometry (gradient flow of the free energy)
\item geometric constraint by the curvature--dimension condition (stability and contractivity)
\end{itemize}
Through this, structure formation and galaxy formation are understood in a unified manner not as separate phenomena, but as a single reaction--transport system in which geometric transport and nonlinear interactions are coupled.
Accordingly, while the present theory starts as a purely measure-theoretic formulation, it can in the future be connected naturally to optimal transport geometry and the theory of gradient flows.

\subsection{Outlook}

In this study, priority was given to the rigor of the theory, and direct correspondence with observations was kept to a minimum. 
However, as established in Section~\ref{sec:measure_theoretic_formulation} and Section~\ref{sec:pushforward_structure}, observation is formulated in a measure-theoretically natural manner as a probability kernel and pushforward operation (Section~\ref{subsec:observation_space_and_kernel}). 
Accordingly, the present theory can be extended directly to statistical estimation for finite-sample data and to the analysis of the time evolution of observables. 

\subsubsection{Future Research Directions}

Important future tasks within this framework include
\begin{itemize}
    \item \textbf{Statistical Estimation:} Construction of a statistical estimation theory incorporating observation kernels and pushforward operations.
    \item \textbf{Data Application:} Application to real data analysis based on the effective two-body theory.
    \item \textbf{Observable Diagnostics:} Examination of the measurability of response coefficients as observables.
    \item \textbf{Cosmological Curvature:} Estimation of the lower curvature bound $K$ and its physical interpretation in a cosmological context.
\end{itemize}
By proceeding in these directions, the measure-theoretic and geometric framework constructed in this paper is expected to facilitate quantitative analysis of galaxy evolution based on observational data. 
A detailed investigation of the observational implications, including the reconstruction of the galaxy manifold from real survey data and concrete tests against cosmological simulations, will be presented in a companion paper \cite{takeuchi2026_merger_practical}.

\subsubsection{Irreversibility and Geometric Limits of the Inverse Problem}

A fundamental challenge identified in this framework is the \emph{intrinsic irreversibility of the pushforward operation} $\Phi_{\#}$ (or the observation kernel $K$). 
As demonstrated in Proposition~\ref{prop:pushforward_noninvertible}, the pushforward is generally non-invertible, implying that the observation process necessarily involves information compression. 
In the context of the galaxy manifold, this non-invertibility defines the geometric limit of the inverse problem: the task of reconstructing the latent evolutionary path $\nu_t$ from the observed measure $\mu_t$ (Section~\ref{sec:pushforward_structure}).

The projection from the latent state space $\mathcal{M}$ to the observation space $\mathcal{X}$ can be viewed as a mapping into equivalence classes where distinct physical states become indistinguishable. 
Consequently, multiple latent trajectories on the galaxy manifold may project onto the same observed sequence in $\mathcal{P}(\mathcal{X})$. 
This geometric degeneracy suggests that data-driven approaches face a fundamental ``information horizon.'' 

Specifically, if the contractivity induced by the ${\rm CD}(K, \infty)$ condition leads to a rapid convergence of latent measures [cite: 1075--1081], the loss of initial state information is further compounded by the observational compression. 
Understanding the structure of this information loss is crucial for establishing the reliability of machine-learning-based reconstructions of cosmic history. 
Future work should focus on characterizing the fiber structure of the pushforward maps to identify which physical degrees of freedom are robustly recoverable. 
This provides not only a warning against over-interpreting projected data but also a rigorous guideline for designing next-generation surveys that minimize such geometric degeneracies.

\acknowledgments

We would like to express our deep gratitude to Shiro Ikeda, Kenji Fukumizu, and Satoshi Kuriki for valuable discussions and insightful comments on this research.

This work was supported by JSPS Grant-in-Aid in Scientific Research (24H00247), and by the Joint Research program of the Institute of Statistical Mathematics (General Research 2) entitled ``Machine-Learning Cosmogony: From Structure Formation to Galaxy Evolution.''



\bibliographystyle{JHEP}
\bibliography{manifold_learning}

\appendix

\section{Foundations of measure theory and probability measure spaces}
\label{app:measure_theory}

\subsection{Polish spaces and probability measures}

\begin{definition}
A metric space $(\mathcal{M},d)$ is called a Polish space if it is complete and separable.
\end{definition}
\medskip
\noindent
In this paper, we assume that the state space $\mathcal{M}$ is a Polish space. Under this assumption, the Borel $\sigma$-algebra $\mathcal{B}(\mathcal{M})$ is naturally defined, and standard results on limits of measures and compactness properties of probability measures can be applied. In particular, it is important that, by Prokhorov's theorem, relative compactness of a family of probability measures is equivalent to tightness.

\begin{definition}
A probability measure on $\mathcal{M}$ is a measure $\mu$ on $\mathcal{B}(\mathcal{M})$ satisfying $\mu(\mathcal{M})=1$.
\end{definition}
\medskip
\noindent
We denote by $\mathcal{P}(\mathcal{M})$ the set of all probability measures. By the Polish property, $\mathcal{P}(\mathcal{M})$ has good topological properties under the topology of weak convergence.

\subsection{Weak convergence and test-function representation}

\begin{definition}
A sequence of probability measures $\{\mu_n\}$ is said to converge weakly to $\mu$ if, for any bounded continuous function $f\in C_b(\mathcal{M})$,
\begin{align}
	\lim_{n\to\infty}\int_{\mathcal{M}}f(z)\,\mu_n(\pd z)=\int_{\mathcal{M}}f(z)\,\mu(\pd z)
\end{align}
holds. In this case, we write $\mu_n\Rightarrow\mu$.
\end{definition}
\medskip
\noindent
This definition characterizes the convergence of measures through duality with functions, and is applicable to general situations without assuming the existence of densities. In particular, the measure evolution treated in this paper is understood as continuity with respect to this weak topology.
Furthermore, by the Portmanteau theorem, this convergence is equivalent to the limit conditions of measures for open sets and closed sets.

\subsection{Probability measure space and second moment}

For the probability measure space $\mathcal{P}(\mathcal{M})$, we impose a moment condition in order to introduce a distance structure.
\begin{definition}
For a reference point $z_0\in\mathcal{M}$,
\begin{align}
	\int_{\mathcal{M}}d(z,z_0)^2\,\mu(\pd z)<\infty
\end{align}
defines the set of probability measures denoted by $\mathcal{P}_2(\mathcal{M})$.
\end{definition}
\medskip
\noindent
In a Polish space, this condition does not depend on the choice of the reference point. The space $\mathcal{P}_2(\mathcal{M})$ serves as the natural domain for the Wasserstein distance introduced later.

\section{Foundations of transport geometry}
\label{app:ot}

\subsection{Wasserstein distance and coupling}

\begin{definition}
For $\mu,\nu\in\mathcal{P}_2(\mathcal{M})$, consider the set of couplings
\begin{align}
	\Pi(\mu,\nu)=\{\pi\in\mathcal{P}(\mathcal{M}\times\mathcal{M})\mid \pi(\cdot\times\mathcal{M})=\mu,\ \pi(\mathcal{M}\times\cdot)=\nu\}
\end{align}
Then the quadratic Wasserstein distance is defined by
\begin{align}
	W_2^2(\mu,\nu)=\inf_{\pi\in\Pi(\mu,\nu)}\int_{\mathcal{M}\times\mathcal{M}}d(z,z')^2\,\pi(\pd z,\pd z')
\end{align}
\end{definition}
\medskip
\noindent
This distance is interpreted as the optimal transport cost between measures. 
$(\mathcal{P}_2(\mathcal{M}),W_2)$ is a complete metric space, and provides the basis for treating time evolution of measures geometrically.
When an optimal transport plan exists for any $\mu,\nu\in\mathcal{P}_2(\mathcal{M})$, a curve of measures $\{\nu_t\}_{t\in[0,1]}$ is defined by interpolation. 
This curve is called a geodesic with respect to the Wasserstein distance.
Furthermore, the limit of the evolution defined by a minimization problem as a time discretization of an energy functional is called the Wasserstein gradient flow, and provides a variational formulation of measure evolution.

\subsection{Pushforward measures}

\begin{definition}
For a measurable map $\Phi:\mathcal{M}\to\mathcal{X}$ and a measure $\mu\in\mathcal{P}(\mathcal{M})$, the pushforward measure $\Phi_{\#}\mu$ is defined by
\begin{align}
	\Phi_{\#}\mu(B)=\mu(\Phi^{-1}(B))
\end{align}
\end{definition}
\medskip
\noindent
This operation is a fundamental method for transferring measures to another space. For any bounded measurable function $f$,
\begin{align}
	\int_{\mathcal{X}}f(x)\,(\Phi_{\#}\mu)(\pd x)=\int_{\mathcal{M}}f(\Phi(z))\,\mu(\pd z)
\end{align}
holds. 
In this paper, the observational measure is defined through this operation.

Moreover, for measurable maps $\Phi,\Psi$, the pushforward satisfies
\begin{align}
(\Psi\circ\Phi)_{\#}=\Psi_{\#}\circ\Phi_{\#}
\end{align}

\subsection{Probability kernels}

\begin{definition}
For measurable spaces $(\mathcal{M},\mathcal{F})$ and $(\mathcal{X},\mathcal{B})$, a map $K:\mathcal{M}\times\mathcal{B}\to[0,1]$ is called a probability kernel if, for any $z$, $B\mapsto K(B\mid z)$ is a probability measure, and for any $B$, $z\mapsto K(B\mid z)$ is measurable.
\end{definition}
\medskip
\noindent
A probability kernel is a generalization of a deterministic map, and describes observational processes including measurement errors and selection effects. For a measure $\nu$,
\begin{align}
	\mu(B)=\int_{\mathcal{M}}K(B\mid z)\,\nu(\pd z)
\end{align}
defines a new measure.

A deterministic map $\Phi$ is written as a special case
\begin{align}
K(B\mid z)=\delta_{\Phi(z)}(B)
\end{align}

\section{Definition of the H-Function and the CD$(K,\infty)$ condition}
\label{app:cd}

For a reference measure $\mathfrak{m}$, define the relative H-function for an absolutely continuous measure $\nu=\rho\mathfrak{m}$ by
\begin{align}
	\mathcal{H}(\nu)=\int_{\mathcal{M}}\rho\ln\rho\,\mathfrak{m}(\pd z)
\end{align}
\begin{definition}
A metric measure space $(\mathcal{M},d,\mathfrak{m})$ is said to satisfy the CD$(K,\infty)$ condition if, for any Wasserstein geodesic $\nu_t$ connecting $\nu_0,\nu_1\in\mathcal{P}_2(\mathcal{M})$,
\begin{align}
	\mathcal{H}(\nu_t)\le(1-t)\mathcal{H}(\nu_0)+t\mathcal{H}(\nu_1)-\frac{K}{2}t(1-t)W_2^2(\nu_0,\nu_1)
\end{align}
holds.
\end{definition}
\medskip
\noindent
This condition provides a lower bound of the effective Ricci curvature in the space of measures. 
It expresses the geodesic convexity of H-function and leads to dynamical properties such as contractivity and stability of measure evolution.
In particular, for the gradient flows of H-function, the contractivity
\begin{align}
	W_2(\nu_t,\tilde{\nu}_t)\le e^{-Kt}W_2(\nu_0,\tilde{\nu}_0)
\end{align}
holds, guaranteeing the stability of the evolution. In this paper, we analyze measure evolution under this geometric constraint.

\end{document}